\documentclass[iop]{emulateapj}

\usepackage{amsmath} 

\begin{document} 

\title{Global Simulations of the Interaction of Microquasar Jets with a Stellar wind in High-Mass X-ray Binaries}
\author{
Yoon, D.\altaffilmark{1} \&
Heinz, S.\altaffilmark{1}
}

\slugcomment{Accepted to ApJ on January 6, 2015}
\altaffiltext{1}{Department of Astronomy, University of Wisconsin-Madison, Madison, WI, USA}

\email{yoon@astro.wisc.edu} 

\begin{abstract} 
  Jets powered by high-mass X-ray binaries must traverse the powerful wind of the companion
  star. We present the first global 3D simulations of jet-wind interaction in high-mass X-ray
  binaries. We show that the wind momentum flux intercepted by the jet can lead to significant
  bending of the jet and that jets propagating through a spherical wind will be bent to an
  asymptotic angle $\psi_{\infty}$.  We derive simple expressions for $\psi_{\infty}$ as a
  function of jet power and wind thrust. For known wind
  parameters, measurements of $\psi_{\infty}$ can be used to constrain the jet power. In the
  case of Cygnus X-1, the lack of jet precession as a function of orbital phase observed by the
  VLBA can be used to put a lower limit on the jet power of $L_{\rm jet} \gtrsim 10^{36}\,{\rm
  ergs\,s^{-1}}$. We further discuss the case where the initial jet is inclined relative to the
  binary orbital axis. We also analyze the case of Cygnus X-3 and show that jet bending is likely
  negligible unless the jet is significantly less powerful or much wider than currently thought.
  Our numerical investigation is limited to isotropic stellar winds. We discuss the possible
  effect of wind clumping on jet-wind interaction, which are likely significant, but argue that
  our limits on jet power for Cygnus X-1 are likely unaffected by clumping unless the global
  wind mass loss rate is orders of magnitude below the commonly assumed range for Cyg X-1.
\end{abstract}

\section{Introduction}

X-ray binaries consist of a compact object, such as a neutron star \citep[{\it e.g.}, Sco
X-1,][]{Hjellming_90} or a black hole ({\it e.g.}  Cygnus X-1, \citealt{Orosz_11}; Cygnus X-3,
\citealt{Zdziarski_13}), accreting mass from a companion. In certain spectral states, the accretion
flow near the compact object generates powerful, collimated jets \citep{Mirabel_99, Gallo_05}.
These jets appear remarkably similar to the jets produced by supermassive black holes (SMBH) in
active galactic nuclei (AGN) in morphology, spectral properties, and energetics (when set in
relation to the overall energy released by accretion).

The mostly featureless non-thermal spectra of relativistic jets limit quantitative analysis of jet
properties to relatively coarse estimates.  However, the propagation of jets and their interaction
with the environment offers a very powerful way to study jet properties that complements direct
studies of jets themselves.  The observations of cavities in galaxy clusters by the {\em Chandra}
X-ray Telescope offer an example of how jet-environment interactions can be used to constrain the
properties of large samples of AGN jets \citep[e.g.][and references therein]{McNamara_07}.

A sub-class in the study of jet-environment interactions involves cases where the accreting black
hole launching the jets is moving relative to the surrounding medium.  In AGN, when the black hole
is moving with considerable speed with respect to the environment, the radio emitting jets are swept
backward by ram pressure, generating a bow shock ahead of the moving black hole.  AGN jets that show
bent morphology are often called ``bent doubles'' or tailed radio sources \citep{Begelman_79,
Freeland_11, Morsony_13}.

It has been suggested that a subset of low mass x-ray binaries (LMXBs) that move through the ISM at
high speed due to the kick velocity the LMXB received in the supernova explosion should also exhibit
bow shocks and trailing neck structures \citep{Heinz_08, Wiersema_09, Yoon_11}, very similar to the
case of bent-double AGN sources.

However, an aspect that makes jet propagation in X-ray binaries fundamentally different from
AGN jets is the presence of the companion star. The early type companion stars of high mass
X-ray Binaries (HMXBs) drive powerful winds in the vicinity of the compact object, which
are often the source of accretion in these objects. The wind mass loss rates of the OB-type
donor star can be substantial: ${\rm \dot{M}_{\rm wind}} \sim 10^{-7}-10^{-5}\, \rm M_{\odot}
\, yr^{-1}$.  The stellar wind dominates the HMXBs' circum-binary environment compared to any
winds launched by the accretion flow.  In this work, we study the dynamics and evolution of
the jets affected by spherical stellar winds from OB-type donor stars in HMXBs.  A subset of
HMXB systems consist of a Wolf-Rayet star in orbit with a black hole \citep[{\it e.g.} Cygnus
X-3][]{Marti_01,Zdziarski_13}; the mass-loss rate from Wolf-Rayet stars is even higher than
that from OB-type stars. Moreover, the Wolf-Rayet star in Cygnus X-3 is tidally locked to the
4.8-h orbital period, resulting in a considerable equatorial enhancement of the mass-loss rate
\citep{Kerkwijk_93}. An investigation of non-spherical and highly-flattened winds is beyond
the scope of this study. While we briefly discuss the case of Cygnus X-3 under the assumption
of a spherical wind, we plan to discuss the effects of non-spherical winds in a future paper.

HMXB winds are not simple. For example, they are likely clumpy \citep{Owocki_88, Oskinova_12}, like
winds generated by other high-mass stars. The interaction of a microquasar jet with clumpy wind
medium has been studied in 3D simulations by \citet{Perucho_12}, which we will refer to as P12
hereafter, who show that clumping can significantly increase jet disruption.  Moreover, even on
average, they are not spherically symmetric due to the gravitational focusing by the compact object
and Coriolis and centrifugal effects due to orbital motion \citep{Friend_82, Miller_05, Hadrava_12}.
In addition, the wind can be ionized by X-rays from the accretion flow, reducing or eliminating line
driving and stalling the wind \citep{Gies_08}. If the X-ray flux is high enough to excite the outer
layer of the star, a thermally driven wind may replace the quenched radiatively driven wind. At the
current time, it is not clear whether the illuminated side of the wind would suffer from the same
line-driving instability that generates clumps in regular massive star winds.

In this pilot paper, we will neglect some of the more poorly understood complications likely present
in binary winds and instead treat the wind as a radiatively driven wind that is isotropic at the
surface of the companion \citep[following, e.g.][]{Castor_75}.  This will allow us to isolate the
fundamental differences in jet propagation in the presence of a wind compared to jets propagating
into uniform medium. We will discuss the effects of some of the likely complications in
\S\ref{sec:caveats}.  The likely most important caveat is the potential clumpiness of the wind; in
that sense, the global simulations presented below should be considered a complementary approach to
the detailed 3D jet-clump simulations presented in P12.

In \S 2, we present the numerical setup and the code used in our parameter study of jet-wind
interaction. In \S 3 we discuss the results of the simulations. In \S 4, we compare the numerical
results with analytic expressions derived for the limiting case of small deflection angles and apply
our model to the HMXBs Cygnus X-1 and Cygnus X-3.  In \S 5 we summarize our results.

\section{Technical Description}

\subsection{The FLASH Code}

Simulations were performed with the FLASH 3.3 hydrodynamics code \citep{Fryxell_00}, which is a
Message Passing Interface(MPI)-parallelized, modular, block-structured adaptive mesh refinement
code.  We employ the non-relativistic unsplit mesh solver, which solves the Riemann problem
using an unsplit staggered mesh scheme on a three dimensional Cartesian grid \citep{Lee_09}.

\subsection{The Wind and Jet Nozzles}

Both the jet and the wind injection are modeled as inflow-boundary conditions on an interior portion
of the grid (we will refer to the regions excluded from the hydro-dynamic integration and instead
treated as an interior boundary as ``nozzles'' following \citealt{Heinz_06b}).

The jet nozzle has a cylindrical shape with inflow boundary conditions at the surface, injecting
a bipolar outflow with a prescribed energy, mass, and momentum flux to match the parameters we
choose for the jet.  For reasons of numerical stability, we inject a slow lateral outflow from
the side walls of the cylinder with negligible mass and energy flux in order to avoid complete
evacuation of zones adjacent to the nozzle due to the large velocity divergence along the jet axis.

The stellar wind nozzle is modeled as a spherical boundary with inflow boundary conditions matching
the desired wind parameters.  We evolve the simulation with only the stellar wind present for one
full orbital period to establish a stable, self-consistent wind profile before switching on the jet
nozzle.

The equation of state is assumed to be adiabatic, tracking two separate phases of the fluid (each
represented by a separate passive tracer fluid to distinguish wind and jet fluids during the
computation and in post-processing). The wind gas is assumed to be a monatomic ideal gas with an
adiabatic index of $\gamma = 5/3$. The internal composition of jets is currently unknown, however,
it is reasonable to assume that they are strongly magnetized and that a sizeable fraction of
their internal energy is carried by relativistic electrons, given the observed synchrotron
radiation.  We investigate jets composed of fluids both with relativistic equation of state with
$\gamma = 4/3$ and with a cold-gas equation of state with $\gamma = 5/3$ and present results in
terms of an unspecified value of $\gamma$ wherever possible.

Most simulations were performed using $\gamma=4/3$, representing the equation of state for a gas
with relativistic internal pressure, either from a fully tangled magnetic field \citep{Heinz_00} or
a relativistic component of the gas; since our simulations are sub-relativistic, a relativistic
equation of state implies that the inertial density is dominated by cold particles (e.g., protons).
A non-relativistic equation of state represents a jet with internal pressure dominated by
non-relativistic thermal plasma. As we will show, our results are only moderately sensitive to the
actual value of $\gamma$.

The gravitational fields of the black hole and the companion are modeled as point source potentials.

To allow for direct comparison with the analytic formulae we derive in \S\ref{subsec:analysis}, most
of the simulations presented in this paper do not include wind driving by radiation pressure.
Instead, we assumed an asymptotic wind at the injection at the stellar surface, {\it i.e.}, a wind with
terminal velocity $v_{\infty}$ and at fixed mass flux of ${\rm \dot{M}_{\rm wind}} \sim 10^{-5}
\,{\rm M_{\odot}\,yr^{-1}}$, which is typical for OB stars \citep{Puls_08}.

Wind driving was incorporated in a sub-set of the simulations presented in this paper to test the
sensitivity of our results against the assumption of an asymptotic radial wind. (See \S\ref{subsec:Accel}). 
For the radiatively driven wind, the initial velocity field follows the so-called $\beta$-law
\begin{equation} \label{eqn:v_beta} v(r) = v_{\infty}
  (1-r_{0}/r)^{\beta},
\end{equation}
where $v_{\infty}=2,500\,\rm km\,s^{-1}$ is the wind terminal speed [chosen to match the wind
parameters of typical OB-type stars \citep{Puls_08}] and $r_{0}=R_{*}x$, where $R_{*}$ is the
radius of the star and $x$ is applied to avoid zero velocity and infinite density on the surface
of the star.  We set the stellar radius $R_{*} = 1.4\times10^{12}\,{\rm cm}$. The value of $x$
can be expressed as $x=\left[ 1 - \left( v_{*}/v_{\infty} \right)^{1/\beta} \right]=0.99$,
where the wind surface velocity, $v_{*}$, is of the order of $10^{6} {\rm cm\,s^{-1}}$, which
is the sound speed with $T_{\rm eff} \approx 30,000\, K$. However, due to the steep variation of
density and pressure around the surface, there is a limit in performing a numerical calculation
with such a high value of $x$.  Alternatively, we set the value of $x=0.95$, which was chosen
to be high enough to maintain the initial wind profile from a typical OB-type star by iterative
1D simulations.  The mass flux of the wind was fixed at the surface of the star (where density
and velocity of the injected wind determine $\rm \dot{M}$ uniquely) and line driving was modeled
using the Sobolev approximation \citep{Castor_74}, such that the total line acceleration results
in $g_{{\rm rad}} \propto \left( \frac{1}{\rho} \frac{dv_{r}}{dr} \right)^{\alpha_{\rm CAK}}$,
where $\alpha_{\rm CAK}$ is the parameter of the CAK model \citep{Castor_75}, typically depending
on the effective temperature of the star. We choose a value of $\alpha_{\rm CAK}=0.64$ in our
model corresponding to a typical OB-type star.

Our simulations are adiabatic and scale-free.  In physical units chosen to approximately match the
wind and binary properties of Cygnus X-1 and allow simulations to be completed within the available
computational resources, the jet velocity was set to be $v_{\rm jet}=3\times10^{9}\,{\rm cm\,s^{-1}}$,
with an initial internal Mach number of ${\mathcal M}_{\rm jet,0}=30$ at the base of the jet (the
``nozzle'').  To explore the dependence on Mach number, we ran a simulation at ${\mathcal M}_{\rm
jet,0}=10$ and otherwise identical parameters compared to our fiducial run with a jet power of
$L_{\rm jet}=10^{36}\,{\rm ergs\,s^{-1}}$. Note that the Mach number ${\mathcal M}_{\rm jet}$ varies
along the jet given the adiabatic behavior of the fluid.

In order to resolve the hydrodynamics at the injection scale with at least 10 cells across the
nozzle, we forced the jet nozzle to be at maximum refinement, resulting in an effective resolution
of $4.7\times 10^{9}\,{\rm cm} = 1.6\times 10^{-3} \,a$, compared to an orbital separation of $a
\approx 3\times 10^{12}\,{\rm cm}$ for Cygnus X-1 \citep{Gies_82}. The radius of the jet nozzle is
about $2.5 \times 10^{10}\,{\rm cm}$ and the full box size of the simulation is about $4 \times
10^{13}\,{\rm cm}$, centered on the center of mass of the binary.  

We varied the jet power to span the range $L_{\rm jet} \approx 10^{35},\,10^{36},\,10^{37}\,{\rm
ergs\,s^{-1}}$, comparable to the range of uncertainty in the jet power of Cygnus X-1, $9 \times
10^{35} - 10^{37}\,{\rm ergs\,s^{-1}}$ \citep{Gallo_05,Russell_07}.

For the bulk of our simulations, the jet was injected in a direction perpendicular to the orbital
plane, {\it i.e.}, along the z-axis of our grid.  We also investigated off-axis jets with angles of 30$^{\circ}$,
60$^{\circ}$, 75$^{\circ}$ relative to the orbital axis of the system, inclined towards the binary companion
(inclination angles perpendicular to the orbital separation vector would not result in any change in
the simulation, given that simulations only cover a small fraction of the binary period once the jet
is switched on). The detailed model parameters of our different runs are described in table
\ref{table:model}.

Note also that in cases where the jet is oriented perpendicular to the binary separation vector
$\vec{a}$, the simulations have mirror-symmetry about $\vec{a}$; in these cases, in order to reduce
the need for computational resources, we only simulate the upper hemisphere at full resolution, but
include both hemispheres to avoid spurious boundary effects near the orbital plane. Results in those
cases are quoted for the high-resolution half of the simulation.

\subsection{Orbital Motion}

The binary parameters were set loosely approximate the parameters for of Cygnus X-1. For
simplicity, we set the mass of the black hole and the star to be 10 and 20 $\rm M_{\odot}$, respectively,
and the separation between them is set to be $3\times 10^{12}\,{\rm cm}$, which gives
an orbital period of 5.8 days for Cygnus X-1, compared to the observed value of 5.6 days
\citep{Brocksopp_99,Pooley_99}

As we show below, the jet propagation time across one binary separation (1.7 minutes) and the time
required for a quasi-stationary bent jet solution to be established (approximately 10 hours) is much
shorter than the orbital time.  For numerical simplicity and to allow direct comparison with the
analytic formulae presented in \S\ref{subsec:analysis}, we neglected orbital rotation in most of our
simulations, keeping the two nozzles stationary in our cartesian grid.  As a first order
approximation, this is justified because the orbital velocity is only of order 20\% of the wind
velocity, and thus orbital effects on the wind ram pressure introduce corrections of the order of
only 5\%.

In order to verify that the effects of orbital motion on the gross dynamics of jet propagation are
small, we ran a sub-set of the simulations including orbital rotation, presented in
\S\ref{subsec:analysis}.  In this case, the nozzles (star and jet) move along their orbital
trajectories in the x-y plane ({\it i.e.}, we simulated the orbit in a fixed, non-rotating frame, which
eliminates the need to introduce terms for Coriolis and centrifugal forces into the solver).  The
coordinate origin was set to be at the center of mass.

In the rotating case, we assumed that the star is co-rotating with the orbit, such that the outflow
velocity at the stellar surface is given by
\begin{equation}
    \vec{v}_{\rm wind,rot}(\vec{x})=\vec{\omega}\times \vec{x} + \vec{v}_{\rm
    wind,*}\left(\vec{x}-\vec{R_{*}}\right),
    \label{eq:vrot}
\end{equation}
where $\vec{\omega}$ is the orbital angular velocity, $\vec{v}_{\rm wind,*}$ the wind velocity at the
stellar surface calculated from eq.~(\ref{eqn:v_beta}), and $\vec{R_{*}}$ the position of the star.

We ran the simulations for one orbital period before switching on the jet in order to allow the flow
to establish a converged velocity and density profile. After the jet launches, the simulations were
carried out for several hours in real time units, long enough to establish the bow shock and the jet
in a quasi-steady state (see \S\ref{subsec:Jetpropa}).

Finally, we performed a comparison simulation of a jet propagating into a uniform wind with
parameters matching those of our fiducial run at the position of the compact object (referred to
below as UniWind\_E36).

\begin{deluxetable*}{ccccc}
  \tablecolumns{4}
  \tabletypesize{\scriptsize} \tablecaption{Model parameters}
  \tablewidth{0pt} 
  \tablecaption{Model Parameters \label{table:model}}
  \tablehead{ \colhead{Model} &
    \colhead{$L_{\rm jet}\,(\rm ergs\,s^{-1})$} & \colhead{${\mathcal M}_{\rm jet,0}$} &
    \colhead{inclination (degree)} & \colhead{$h_{1}\tablenotemark{b} \, ({\rm cm\,s^{-1}})$}}
  \startdata
  SphWind\_E35      & $10^{35}$ & 30 & 0\tablenotemark{a} & $3\times10^{10}$  \\
  SphWind\_E36      & $10^{36}$ & 30 & 0 & $8\times10^{10}$\\
  SphWind\_E36\_M10   & $10^{36}$ & 10 & 0 & $1\times10^{11}$\\
  SphWind\_E36\_rot & $10^{36}$ & 30 & 0 & $8\times10^{10}$ \\
  SphWind\_E36\_acc & $10^{36}$ & 30 & 0 & $8\times10^{10}$ \\
  SphWind\_E36\_30deg & $10^{36}$ & 30 & 30 & $8\times10^{10}$\\
  SphWind\_E36\_60deg & $10^{36}$ & 30 & 60 & $8\times10^{10}$\\
  SphWind\_E36\_75deg & $10^{36}$ & 30 & 75 & $8\times10^{10}$\\
  SphWind\_E37      & $10^{37}$ & 30 & 0 & $1.5\times10^{11}$\\
  SphWind\_E37\_gam166  & $10^{37}$ & 30 & 0 &  $1.5\times10^{11}$\\
  \tableline UniWind\_E36 & $10^{36}$ & 30 & 0 & $8\times10^{10}$
  \enddata   
  \tablenotetext{a}{$0^\circ$ indicates that the direction of the
    jet is perpendicular to the line between the star and the Black
    hole.}  \tablenotetext{b}{The jet thickness at the re-collimation
    shock, $h_{1}$, is measured naively by checking the variation of
    the jet thickness along the jet from simulation results.}
\end{deluxetable*}

\subsection{Measurement of the Jet Thickness and Propagation Direction}\label{subsec:Jetpropa}

A key variable determining the strength of the jet-wind interaction is the thickness $h$ of the jet
as seen by the wind ({\it i.e.}, the size of the jet perpendicular to both jet and wind velocities). We
measured $h$ as follows.

In post-processing, we identified matter inside a computational cell as jet material if the value of
$ \zeta \equiv \left( v_{z}/v_{\rm jet} \right) J$ exceeded a fixed threshold, where $\left( v_{z}/v_{\rm jet}
\right)$ is the flow velocity normalized by the initial jet speed, which is $3\times 10^9 \, {\rm
cm\, s^{-1}}$ in our standard parametrization, and $J$ is the fractional density of jet tracer fluid
injected at the nozzle.  We found that a choice of $\zeta = 0.1$ successfully identified the jet
material in all cases (see Figure~\ref{fig:jetthick}). The thickness of the jet was measured in the
direction perpendicular to the separation vector and the jet axis, since it is the dimension of the jet
in that direction that determines the amount of wind momentum flux intercepted by the jet.  

When the jet first turns on, it propagates in its initial direction, following the standard
evolution of jet propagation until the expansion of the cocoon becomes slower than the wind
velocity.  As the head of the jet propagates, the accumulated perpendicular momentum flux begins
to bend it away from its initial propagation direction.  We then traced the propagation direction
of the jet fluid to determine the jet trajectory and bending angle.

\section{Results}

\subsection{Jet Bending in Spherical Winds}\label{subsec:evol}

Our simulations confirm the general expectation that a powerful wind from a companion star can
affect the propagation of the jet, and that the ultimate trajectory of the jet depends on the
relative momentum flux in the wind and the jet, as well as the geometry of the system.

\begin{figure*}
  \centering
  \includegraphics[width=\textwidth]{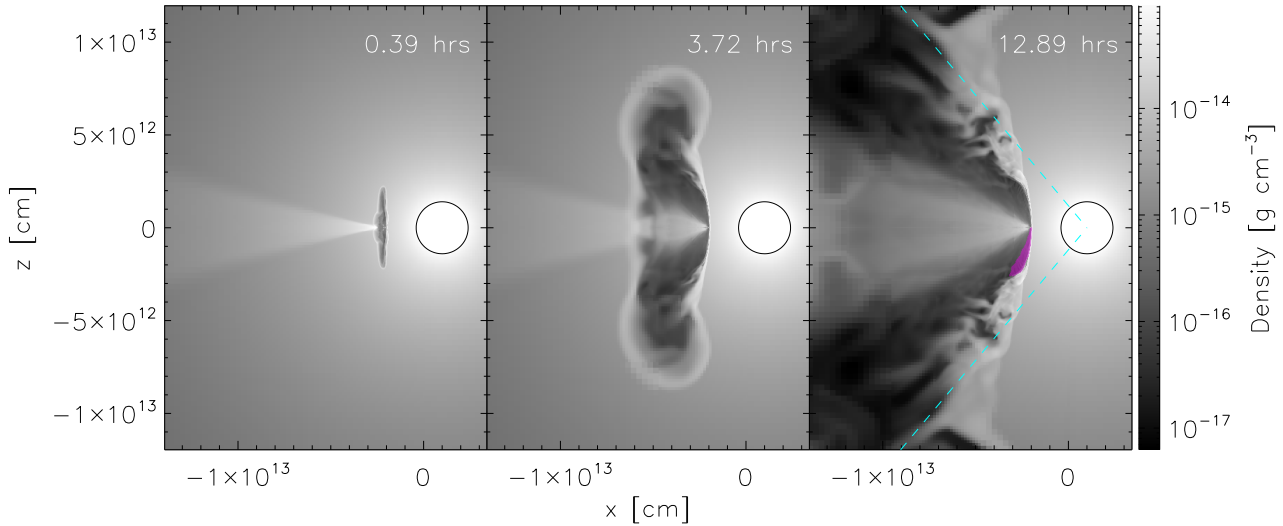}
  \caption{Time sequence of density maps for our fiducial simulation
    SphWind\_E36. The black circle indicates the surface of the companion star. The enhanced density
    in the down stream of equatorial plane to the left of the black hole is due to the
    gravitationally focused wind. The bow shock structure along the jet reaches steady state
    approximately in 12 hours after it launches. The magenta area in the right-most image indicates
    the jet materials (marked only in the lower half of the image), identified by a certain
    threshold (see \S\ref{subsec:recollanal}).  The cyan dashed lines in the right-most panel
    indicate asymptotic lines along which the jets converge, showing that the jet is bent by
    approximately 30$^\circ$ from the initial direction. In the down-stream region, the bow-shocked
    wind passes around the jet and re-collimates in an expansion fan and a (weak) re-collimation
    shock, as expected for super-sonic flow around an object, leaving the post-shock region filled
    with wind gas, visible in the right-most two panels.}
  \label{fig:1e36den}
\end{figure*}

We briefly describe the morphology of the jet-wind interaction and compare it to simulations
of jet bending observed in the interaction of a jet with a uniform medium \citep{Yoon_11}.
Figure~\ref{fig:1e36den} shows snapshots of our fiducial run at a jet power of $L_{\rm
jet}=10^{36}\,{\rm ergs\,s^{-1}}$.

Upon injection into the grid, the jet fluid is generally over-pressured compared to the external
pressure and the ram pressure of the stellar wind (the pressure is fixed by our choice of the jet
power, the jet velocity, the Mach number, and the cross sectional area of the jet nozzle).

We chose this setup to allow the jet to establish a self-consistent, stable structure by letting
the jet reach pressure equilibrium with the bow shock.  During this phase, the jet does not
experience any bending, given that it is strongly over-pressured with respect to the ram pressure
in the stellar wind. We show a typical setup in Figure~\ref{fig:1e36den}, which
displays a density slice through the simulation. Generally, the lack of external confinement
leads to free lateral expansion of the jet with a half-opening angle of order $\alpha_{0} \sim
1/{\mathcal M}_{\rm jet,0}$.  The wind is gravitationally focused by the black hole in the down
stream region, generating the density enhancement along the equatorial plane visible to the left
of the black hole in Figure~\ref{fig:1e36den}. Here and in the following, we place the x-axis along
the orbital separation vector at t=0 and the z-axis along the orbital angular momentum vector.

The free expansion of the jet proceeds until its pressure reaches the pressure behind the bow shock
of the wind, at which point the jet goes through a re-collimation shock and reaches a stable
equilibrium thickness $h_{1}$.  Thus, $h_{1}$ is not directly set as a simulation parameter, but instead
determined by the initial Mach number of the jet.

However, since neither the re-collimation region nor the initial Mach number of the jet are
observable, we will carry out most of the analysis in this paper using the jet thickness $h$ beyond
the re-collimation shock, relating it where possible to observables like the large scale opening
angle of the jet $\alpha_{\rm obs}$.  We will describe the details of the re-collimation region,
which is very close to the black hole compared to the size of the simulation box, in
\S\ref{subsec:recoll}.

The approaching stellar wind material goes through a stationary bow shock as it is forced to
propagate around the jet. Beyond the re-collimation shock, the transverse pressure gradient imparted on the jet by the lateral
momentum flux of the stellar wind then gradually bends the jet fluid away from the companion star,
while the jet thickness $h$ is set by lateral pressure balance with the wind bow shock pressure.

The effects of the radially declining wind density and the changing velocity of the wind as a
function of distance along the jet imprint a qualitatively different asymptotic behavior of the jet
compared to interaction with a uniform medium.  Because the density declines roughly as $r^{-2}$,
where $r$ is the distance from the center of the star, the effect of bending declines with distance,
and most of the bending occurs within about a binary orbital separation from the jet nozzle.

More importantly, jet bending is caused only by transverse momentum flux, which depends on the
angle $\vartheta = \sin^{-1}{\left( \left| \hat{\vec{v}}_{\rm jet}\times\hat{\vec{v}}_{\rm
wind} \right| \right)}$ between the local jet velocity and wind velocity through
$\sin^2\left({\vartheta}\right)$. Because $\vartheta$ decreases with increasing $r$, the
amount of transverse momentum flux also decreases with $r$.  Asymptotically, any initially
large $\vartheta$ will tend to zero (even in the absence of any jet bending) and the amount of
lateral momentum flux across the jet decreases strongly with distance.  At infinity, jet and wind
travel parallel to each other at some asymptotic angle $\psi_{\infty}$ relative to the initial
jet direction. In contrast, in a uniform wind, $\vartheta$ does not decrease with distance from
the nozzle, and both jets must eventually be bent such that the jet plasma asymptotically flows
parallel to the wind direction.

\begin{figure}[!htbp]
  \includegraphics[width=0.51\textwidth]{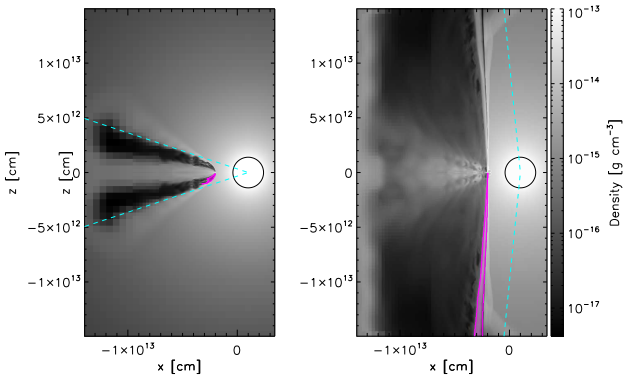}
  \caption{Density maps for the case of SphWind\_E35 (left
    panel) and SphWind\_E37 (right panel) after the steady state bow shock structure
    has been established. While the jet for SphWind\_E35 is disrupted within a short distance from
    its injection, the jet for SphWind\_37 is steadily maintained (marked in magenta color). The
    cyan dashed lines indicate that the jet bending angles are 65$^\circ$ and 8$^\circ$ for
    SphWind\_E35 and SphWind\_E37, respectively.  The vertical black thin trajectory in right
    panel is the low density area generated by the shear layer between the jet and the bow shock.}
  \label{fig:1e35_37}
\end{figure}

We ran simulations with jet powers of $L_{\rm jet} = 10^{35},\,10^{36},\,10^{37}\,{\rm
ergs\,s^{-1}}$, and the results are shown in Figure~\ref{fig:1e35_37}.  The dashed lines
indicate the converged asymptotic lines towards which the jet is bent by the wind, showing that
the $\psi_{\infty}$ is a strong function of jet power.  In the case of $L_{\rm jet}=10^{35}
\rm \,ergs\,s^{-1}$, the ram pressure by the stellar wind is sufficiently strong to bend
the jet by almost 90$^\circ$, similar to the case of the UniWind\_E36 uniform wind model.
On the other hand, for the highest jet power case in our simulation, $L_{\rm jet}=10^{37}\rm
\,ergs\,s^{-1}$, the bending angle is small. We will discuss the relationship between the jet
kinetic power and the inclination angle in \S\ref{subsec:analysis}.

We generally find that the jet is bent towards its asymptotic bending angle within a time scale of
$\tau_{\rm bend} \sim 0.05 \,\tau_{\rm orbit}$, where $\tau_{\rm orbit}$ is the orbital period; this
is roughly the time for the geometry of the inner jet to reach a steady state.  We measured the
propagation direction and asymptotic bending angle of the jet after it settled into its steady
state.  Note that the short bending time scale $\tau_{\rm bend} << \tau_{\rm orbit}$ implies that
the jet reacts instantaneously to changes in binary orbit, which further implies that the jets must
be precessing on the orbital period of the system if bent by jet-wind interaction.

\begin{figure}[!htbp]
  \includegraphics[width=0.51\textwidth]{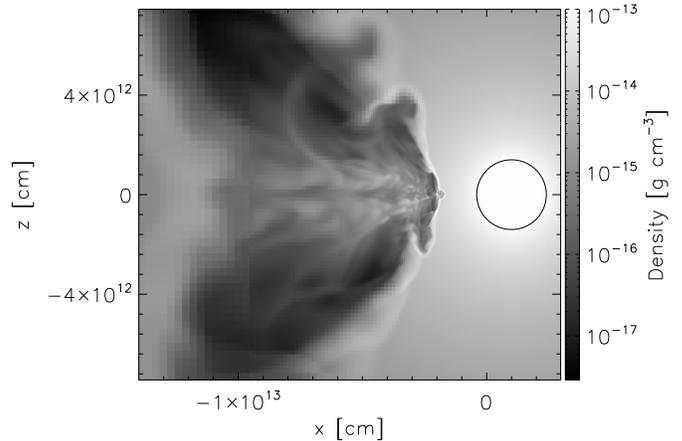}
  \caption{Density map in the case of lower Mach number, 
    ${\mathcal M}_{\rm jet,0}$ = 10 (SphWind\_E36\_M10).}
  \label{fig:1e36M10}
\end{figure}

Figure~\ref{fig:1e36M10} explores the dependence of our simulations on the initial internal Mach
number of the jets, with ${\mathcal M}_{\rm jet,0}$ set to one third of our fiducial value.  With
otherwise identical parameters, a smaller Mach number implies larger thermal energy relative to the
total energy of jet.  The increased thermal pressure leads to a larger initial opening angle of the
jet, which in turn results in an increase in transverse momentum transfer and jet bending.  The
increased surface area and decreased Mach number also increase the incidence of Kelvin-Helmholtz
instability and earlier onset of jet disruption.

\subsection{The Re-collimation Shock}\label{subsec:recoll}

\begin{figure}[!htbp]
  \includegraphics[width=0.51\textwidth]{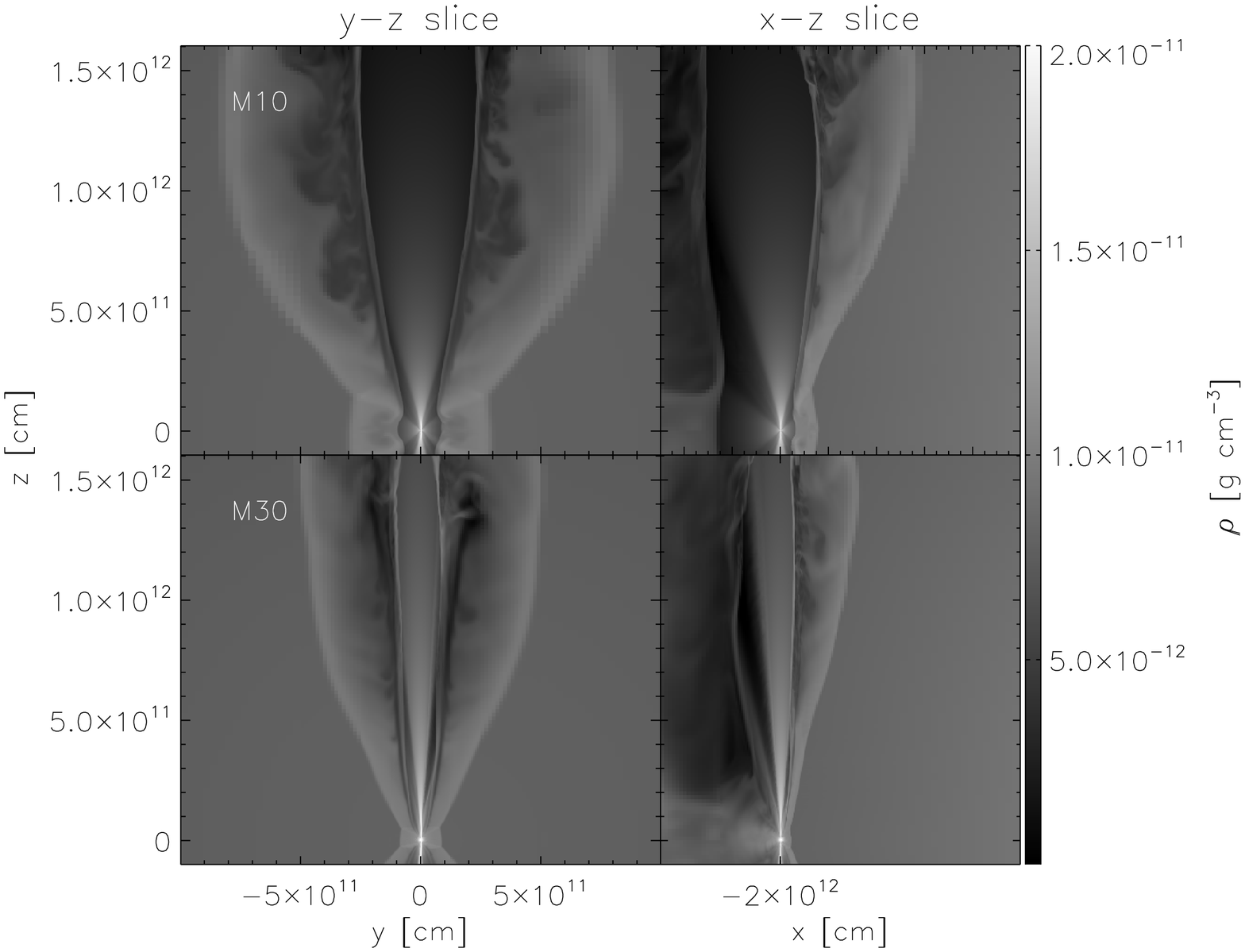}
  \caption{Density map for re-collimating jets in ${\mathcal
      M}_{\rm jet,0}=10$ (upper panels), and ${\mathcal M}_{\rm
      jet,0}=30$ (lower panels). The jet is re-collimated in the y-z plane while bending occurs in the x-z plane.}
  \label{fig:recoll} 
\end{figure}

In order to investigate the physics of jet re-collimation by the wind, we carried out a set
of test simulations with significantly increased resolution, restricted to a shorter duration.
We performed the tests with two jet Mach numbers, ${\mathcal M}_{\rm jet,0}=10, 30$. Snapshots
of the re-collimation region are shown in Figure~\ref{fig:recoll}. We measured
the jet thickness along the y-axis (y-z slice) for the analysis below because the effective
cross section of the wind momentum flux captured by the jet depends only on the width of the
jet in y-direction. Jet bending is facilitated by the pressure gradient in the x-z plane,
where only the leading edge of the jet is subject to the increased pressure behind the bow shock.

In our simulations, the jet is initially freely expanding.  Acceleration of the lateral expansion
will become inefficient once the lateral motion itself becomes supersonic.  This sets the
characteristic semi-opening angle $\alpha_{0}$ of such a supersonic ``fan'' simply as
\begin{equation}
  \alpha_{0}\sim \frac{1}{\mathcal M}_{\rm jet,0} = \frac{c_{\rm s,0}}{v_{\rm jet}} =
  \sqrt{\frac{\gamma P_{0}}{\rho_{0}}}\frac{1}{v_{\rm jet}} .
\end{equation}

\begin{figure*}[!htbp]
  \begin{center}$
    \leavevmode
    \begin{array}{cc}
        \epsfxsize=7cm\epsfbox{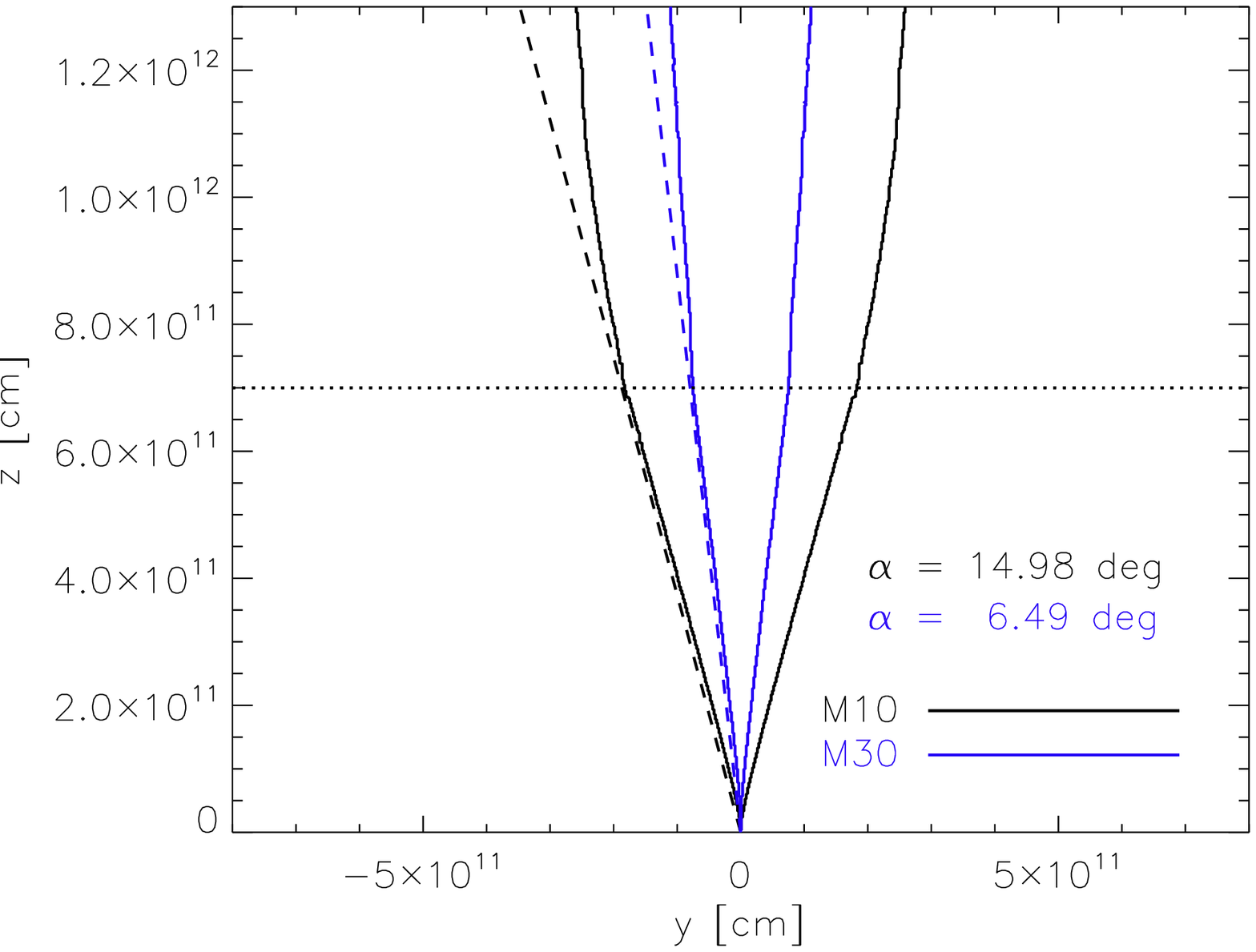} &
        \epsfxsize=7cm\epsfbox{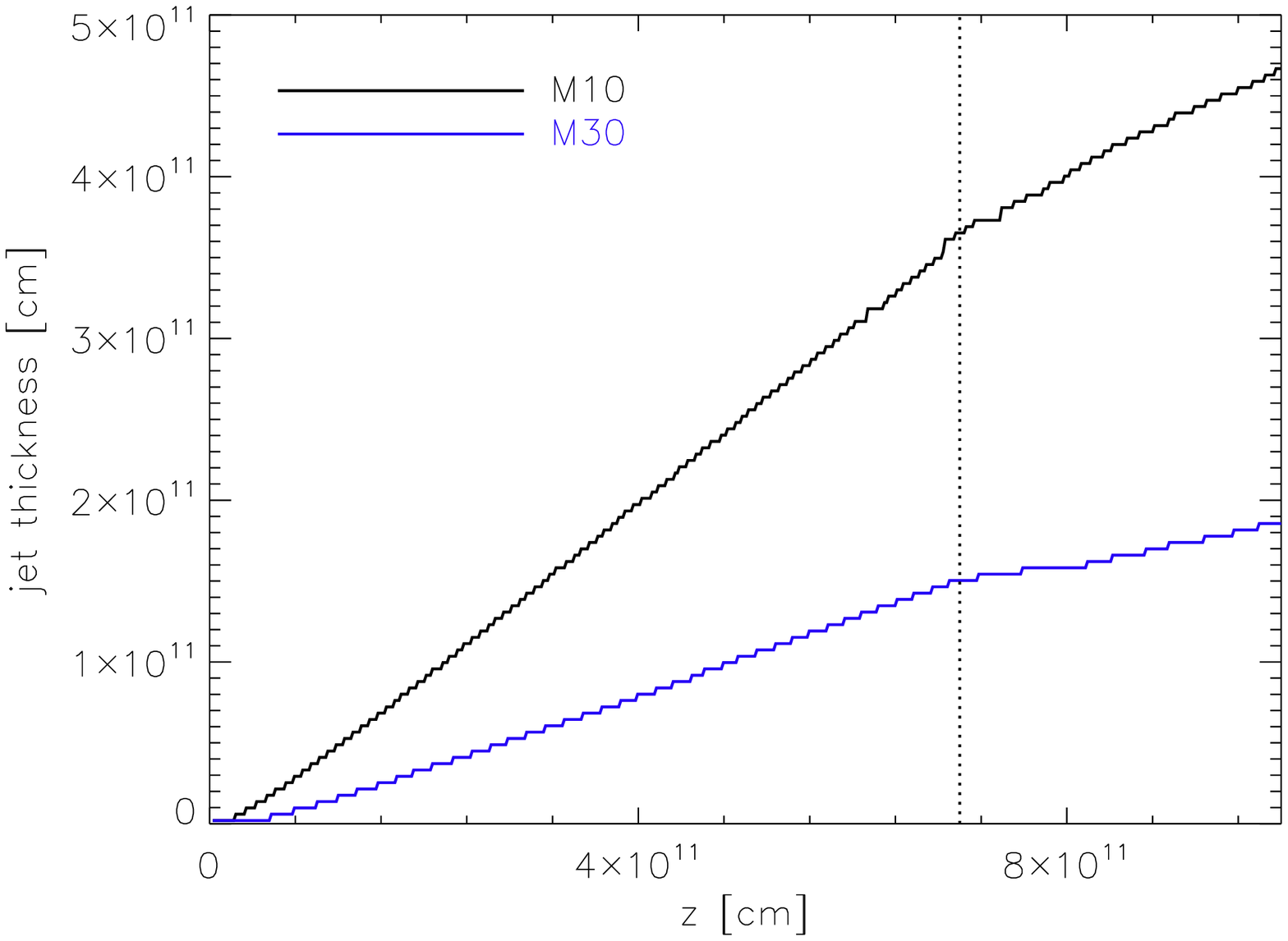}
    \end{array}$ 
  \end{center} 
  \caption{Left panel: Identified jets from simulation for the case of ${\mathcal M}_{\rm jet,0}=10, 30$. 
    Right panel: Measured jet thickness along jet. The dotted lines indicate the location of 
    re-collimation shock.}
  \label{fig:recolljet} 
\end{figure*}

In Figure~\ref{fig:recolljet} we plot the measured jet thickness $h$ as a function of height $z$.  As
expected, the jet with the initially higher Mach number has a narrower opening angle.

From the numerical experiment, the initial half-opening angle $\alpha_{0}$ of the jet is roughly
\begin{equation}
    \alpha_{0} \sim \frac{3}{{\mathcal M}_{\rm jet,0}}
\end{equation}
slightly larger than the simplistic estimate $\alpha_{0} \sim 1/{\mathcal M}_{\rm jet,0}$.

Once the conical expansion has been established, the lateral ram pressure $P_{\rm jet,ram,\perp}$ of
the jet
\begin{equation}\label{eq:P_jetRam}
  P_{\rm jet,ram,\perp} = \rho_{\rm jet}\sin^{2}{\alpha_{0}}
  \,v_{\rm jet}^2
  = \rho_{0}\left(\frac{z_{0}}{z}\right)^2\sin^{2}\alpha_{0} \,v_{\rm
    jet}^2
\end{equation}
is always larger than the internal pressure (since the lateral expansion is supersonic).

In terms of the kinetic jet power
\begin{equation}
  L_{\rm jet,kin} = \pi \rho_{\rm jet} v_{\rm
    jet}^3 h^2/4 = \pi \rho_{\rm jet} v_{\rm jet}^3 z^2 \sin^{2}{\alpha_{0}} 
\end{equation}
the lateral ram pressure is
\begin{equation}
  P_{\rm jet,ram,\perp} = \frac{L_{\rm jet,kin}}{\pi z^2 v_{\rm jet}}
\end{equation}
independent of $\sin\alpha_{0}$ and ${\mathcal M}_{\rm jet,0}$.

Lateral expansion will proceed until $P_{\rm jet,ram,\perp}$ drops below the pressure in the wind bow
shock that forms around the jet, $P_{\rm wind,ram}$. At this point, a re-collimation shock must form
in the jet and bring the internal pressure of the jet into equilibrium with the bow shock. We will
denote the location of the re-collimation shock along the jet as $z_{1}$.  For parameters considered
in this paper, $z_{1}$ is always much smaller than the binary separation $a$, so we will assume
that ram pressure of the wind is constant for the discussion of $z_{1}$ and $h_{1}$, so the wind
ram pressure is given by its value in the equatorial plane at the location of the black hole.

In terms of the mass loss rate of the wind\footnote{Given typical properties of OB-type stars, the
mass loss rate is of order $\dot{\rm M}_{\rm OB} \sim 10^{-7}\sim10^{-5}\,\rm M_{\odot}\,yr^{-1}$ with a
terminal velocity of $v_{\rm wind} \approx 2000 - 3000\,\rm km\,s^{-1}$ \citep{Castor_75,Puls_08}.},
${\rm \dot{M}_{\rm wind}}=4\pi r^2 v_{\rm wind} \rho_{\rm wind}=4\pi a^2 v_{\rm wind} \rho_{\rm wind,0}$,
the wind ram pressure at the jet nozzle is then
\begin{equation}\label{eq:P_windRam}
  P_{\rm wind,ram,0} = \rho_{\rm wind,0} v_{\rm wind}^2 =
  \frac{{\rm \dot{M}_{\rm wind}}v_{\rm wind}}{4\pi a^2}
\end{equation}
where, $\rho_{\rm wind,0}$ is the wind density at the footpoint of the jet and $v_{\rm wind}$ is the velocity of the stellar wind, assumed to be constant.

The location of the re-collimation shock $z_{1}$ is given by equating $P_{\rm jet,ram,\perp}=P_{\rm wind,ram,0}$:
\begin{equation}\label{eq:z1}
  z_{1} = a \sqrt{\frac{4\,L_{\rm jet}}{{\rm \dot{M}_{\rm wind}}v_{\rm
        wind}v_{\rm jet}}}
\end{equation}
again independent of ${\mathcal M}_{\rm jet,0}$ and $\alpha_{0}$.

This simple picture is qualitatively confirmed by the high-resolution simulations of
the re-collimation shock: Figure~\ref{fig:recolljet} shows the re-collimation shock at $z \approx 7
\times 10^{11} {\rm cm}$ {\em regardless} of the jet Mach number, consistent within an
error range of 20\% with the analytic solution $z_{1}\approx9\times 10^{11}\, {\rm cm}$ from
eq.~(\ref{eq:z1}) for the simulation parameters ($L_{\rm jet}=10^{37} {\rm ergs\,s^{-1}}$, ${\rm
\dot{M}=10^{-5} M_{\odot}\,yr^{-1}}$, $v_{\rm wind}=2.5\times10^{8}\,{\rm cm}$, $v_{\rm jet}=0.1\,c$,
$a=3\times10^{12}{\rm cm}$).

At the re-collimation shock, the jet will have a thickness $h_{1}$ and will be in pressure
equilibrium. Beyond $z_{1}$, the jet will thus adjust its thickness $h$ to maintain pressure
equilibrium with the wind.  We will discuss the propagation of the jet in this phase, and the
interaction with the wind that occurs beyond $z_{1}$, in the next section.

\section{Discussion}

As described above, the development of an asymptotic bending angle $\psi_{\infty}$ is expected from
simple considerations of the geometry of the interaction between jet and wind.

A full analytic description of the detailed dynamical evolution of the jet (including the onset of
dynamical instabilities such as Kelvin-Helmholtz instability) is beyond the scope of this paper.
Therefore, for the following analysis, we make the assumption that the bending angle $\psi$ of the
jet is small (implying relatively weak interaction).

Because non-linear effects (like dynamical instabilities) tend to increase the cross section of the
jet, they will tend to increase the bending angle due to the large net transverse momentum
intercepted by the jet. In that sense, the relations derived below will be lower limits on the
actual bending angle. This is borne out by our simulations (See Figure~\ref{fig:jet_trace} below).

We will further assume that the wind is asymptotic, {\it i.e.}, has reached constant velocity before
interacting with the jet and thus follows a simple $r^{-2}$ density profile.  We will also neglect
effects of orbital motion in the analytic approximations below (justified by the fact that they can
be expected to be about an order of magnitude smaller than the dominant effects, as argued above).

\subsection{The Evolution of the Jet Thickness Beyond the Re-Collimation
Shock}\label{subsec:recollanal}

We will assume that the jet is in pressure equilibrium with the ram pressure of the wind,
\begin{eqnarray}\label{eq:Pram_con}
    P_{\rm jet,ram,\perp} &=& \rho_{\rm wind}(\theta) \left( v_{\rm wind} \cos^{2}{\theta} \right)^2 \nonumber \\
    &=& \rho_{\rm wind,0}v_{\rm wind}^2\cos^{4}{\theta}  \nonumber \\
    &=& \rho_{\rm wind,0}v_{\rm wind}^2\left(\frac{a^2}{a^2 + z^2}\right)^2
\end{eqnarray} 
where we have used $\cos{\theta} = a/\sqrt{a^2 + z^2}$, where $\theta$ is the angle between
the orbital separation vector $\vec{a}$ connecting the star and the black hole, and the vector
$\vec{r}$ from the star to a given position along the jet.  The jet has an initial jet thickness
$h_{1}$, as discussed above, which we take as a parameter in the following.

\begin{figure}[!htbp]
    \centering \includegraphics[width=0.51\textwidth]{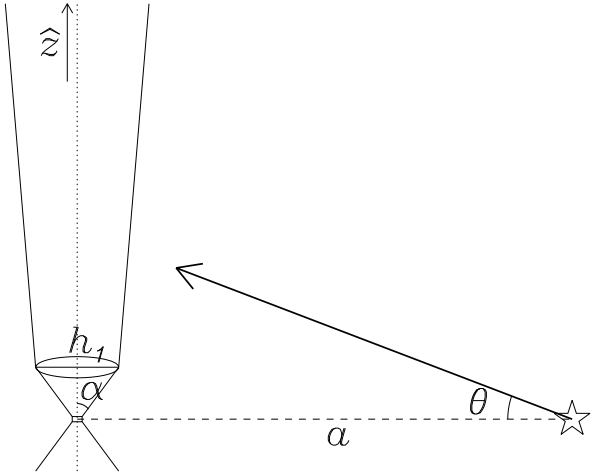} 
    \caption{Schematic figure. $\alpha$ is a jet semi-opening angle,
	$a$ is a separation, $\theta$ is an inclination angle from the orbital plane, and $h_{1}$ is
	the jet thickness at the collimation shock.}
     \label{fig:schem}
\end{figure}

The geometry of the jet and the re-collimation shock and the definitions of the relevant angles
and coordinate axes are sketched in Figure~\ref{fig:schem}.  

Beyond the re-collimation shock, the jet thickness $h$ follows from pressure equilibrium between the
jet and the bow shock:
\begin{eqnarray}\label{eq:pres_equil}
    P_{\rm jet,ram,\perp} &=& \rho_{\rm wind,0} v_{\rm wind}^2  \left(\frac{a^2}{a^2 +
       z^2}\right)^2  \nonumber \\
       &=& P_{\rm jet} =P_{\rm eq,1} \left[ \frac{h(z)}{h_{1}} \right]^{-2 \gamma}
\end{eqnarray} 
where $P_{eq,1}$ and $h_{1}$ are the pressure and the jet thickness at the re-collimation
shock. This sets the jet thickness $h$: 
\begin{equation}\label{eq:thick_height}
    h\left( z \right) = h_{1}\, \left(\frac{a^2}{a^2 + z^2}\right)^{-1/\gamma}
\end{equation}

\begin{figure}[!htbp]
  \centering \includegraphics[width=0.51\textwidth]{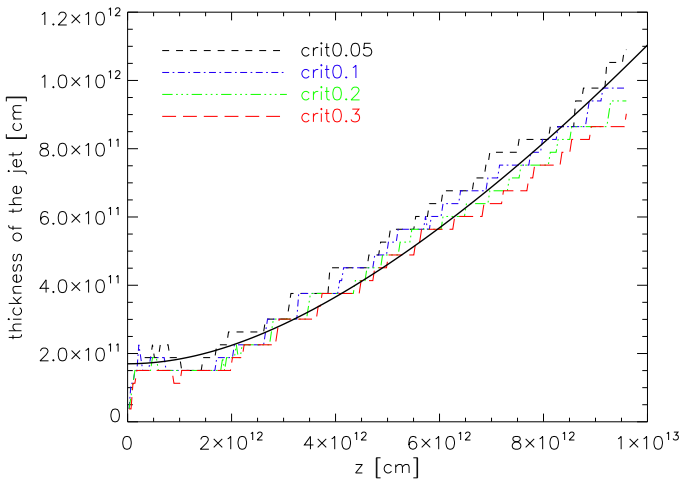} 
  \caption{The thickness of
    the jet in SphWind\_E37 model. The solid
    black line represents the analytic solution from eq.~(\ref{eq:thick_height}) with the parameters
    appropriate for the model of SphWind\_E37, and dashed and dot-dashed lines indicate the
    numerical results for different choices of the threshold used to determine whether a
    computational cell belongs to the jet.}
\label{fig:jetthick} \end{figure}

Figure~\ref{fig:jetthick} shows the measured jet thickness $h$ as a function of $z$ for simulation
SphWind\_E37, compared to the value calculated from eq.~(\ref{eq:thick_height}), for different
choices of the jet threshold (see \S\ref{subsec:Jetpropa}). The figures show good agreement between
the model and the simulation.

Generally, $h_{1}$ will not be measurable. However, we can us eq.~(\ref{eq:thick_height}) to relate
a measured jet thickness (or an upper limit) at large $z$ to the jet thickness at any other $z$,
given values for $\gamma$ and $a$.

\citet{Stirling_01} reported that the VLBA jet of Cygnus X-1 has a semi-opening angle of
$\alpha_{\rm VLBA} = h/2\,z_{\rm VLBA} \lesssim 2^{\circ}$, where $z_{\rm VLBA}$ is the scale
length of the extended jet on which the opening angle is measured; the orbital separation and
the length of the jet are 0.1 mas and 15 mas, respectively. By using eq.~(\ref{eq:thick_height})
with $\gamma=4/3$, the jet thickness at the re-collimation shock can be estimated as $h_{1}
\lesssim 5.7 \times 10^{-3} \,a$.  For a value of $\gamma=5/3$, we find a value of $h_{1}
\lesssim 1.3 \times 10^{-2}\,a$. We will further discuss the constraints on the initial
half-opening angle in the case of the Cygnus X-1 jet in \S\ref{subsec:cygx1}.

\subsection{Jet Bending and the Asymptotic Bending Angle}\label{subsec:analysis}

With an expression for the jet thickness $h$ from eq.~(\ref{eq:thick_height}), we can now discuss
the amount of bending experienced by the jet.  Technically, jet bending occurs because a transverse
pressure gradient exists behind the bow shock that the wind drives around the jet, such that
the external pressure at the leading edge is $P_{\rm 1} \sim P_{\rm bow}$ and the pressure on
the trailing edge of the jet is $P_{\rm 2} \ll P_{\rm 1}$.  Thus, a transverse pressure gradient
exists inside the jet as well, acting to accelerate/bend the jet fluid away from the star.


In deriving an estimate for $\psi_{\infty}$, we will make the simplifying assumption that the
bending angle is small, {\it i.e.}, that the accumulated transverse momentum flux is small
compared to the lateral momentum flux. The reason for this assumption is that the jet will be
dynamically disrupted if the bending angle is large, as the jet-boundary interaction must be
significant in this case.  Our simulations bear out the validity of this assumption at least in
the hydrodynamic case studied here. Similarly, we neglect the momentum transfer from the wind
to the jet in the longitudinal direction, which would lead to acceleration or deceleration by
a small amount.

Based on the properties of observed astrophysical jets (the presence of shocks, the inferred
large kinetic power compared to the minimum internal energy based on the observed synchrotron
intensity), we further assume the jet to be supersonic (large internal Mach number) and ballistic
(no further acceleration beyond the nozzle). This reflects the setup of our simulations.


Under the assumption of constant longitudinal jet velocity and $\mathcal{M}_{\rm jet,0}^2 \gg
1$, the longitudinal jet momentum per unit jet length is conserved and given by
\begin{eqnarray}\label{eq:momJ}
    \Phi_{m,{\rm jet}} &=& \int dA_{\perp} \rho_{\rm jet}v_{\rm jet} \nonumber \\
                       &=& \pi\,r_{\rm jet,0}^2\,\rho_{\rm jet,0}\, v_{\rm jet} \nonumber \\
                       &=& \frac{L_{\rm jet,kin}}{v_{\rm jet}^2}  ,
\end{eqnarray} 
where $dA_{\perp}$ is the area element perpendicular to the initial jet direction, $r_{\rm jet,0}$
and $\rho_{\rm jet,0}$ are the radius and the density of the jet at the nozzle.

The transverse momentum per unit jet length accumulated by the jet can be derived as a function
of $z$:
\begin{eqnarray}
  \label{eq:momW} 
  \Delta \Phi_{m,{\rm wind}} 
 &=& \int_{0}^{t(z)} dt \int_{-h/2}^{h/2} dy
 \int_{-X(y)/2}^{X(y)/2} dx \nabla_{x} P_{\rm bow} \nonumber \\
 & = & \int_{0}^{z} \frac{dz'}{v_{\rm jet}} \int_{-h/2}^{h/2} dy P_{\rm bow} \nonumber \\
 & = & \int_{0}^{z} \frac{dz'}{v_{\rm jet}} h(z') P_{\rm bow}  \nonumber \\
 & = & \int_{0}^{z} \frac{dz'}{v_{\rm jet}}\rho_{\rm wind,0}v_{\rm  wind}^2
 \, h_{1} \left(\frac{a^2}{a^2 + {z'}^2}\right)^{2-1/\gamma}\nonumber \\
 & = & \frac{h_{1} \rho_{\rm wind,0} v_{\rm wind}^2} {v_{\rm jet}} a
 f(z,\gamma) \nonumber \\
 & = & \frac{h_{1}{\rm \dot{M}}_{\rm wind}v_{\rm wind}}
 {4\pi\,a\,v_{\rm jet}} f(z,\gamma)
\end{eqnarray}
where
\begin{equation}
  f(z,\gamma)\equiv \int_{0}^{z/a} dy \left(\frac{1}{1 +
      y^2}\right)^{2 - 1/\gamma}
\end{equation}
which can be expressed as a combination of hypergeometric functions, but is most easily evaluated
numerically.

In the first order (small bending angle) approximation we are making here, the ratio of transverse
to longitudinal momentum is equal to the bending angle ({\it i.e.}, the angle between the local
and the initial velocity vector or tangent vector of the jet as a function of $z$):
\begin{eqnarray}
  \psi(z,\gamma) & = & \frac{v_{\perp}}{v_{\parallel}} = \frac{\Delta
\Phi_{m,{\rm wind}}}{\Phi_{m,{\rm jet}}} \nonumber \\
  & = &
\frac{{\rm \dot{M}}_{\rm wind}v_{\rm wind} v_{\rm jet}\,h_{1}}{4\pi\,a\,L_{\rm
      jet,kin}} f(z,\gamma)
  \label{eq:psi}
\end{eqnarray}

\begin{figure}[!htbp]
   \includegraphics[width=0.51\textwidth]{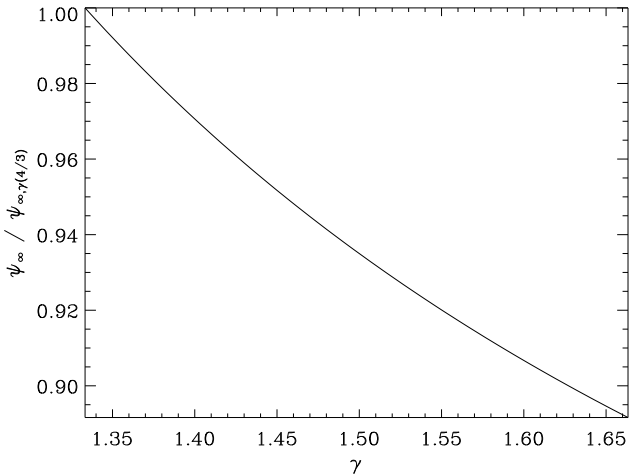}
   \caption{The asymptotic jet bending angle as a function of
     adiabatic index, $\gamma$. The angle is normalized by the one for
     the case of $\gamma=4/3$.}
\label{fig:angGam}
\end{figure}

The asymptotic value for $z \longrightarrow \infty$ can then be evaluated in terms of elementary
Gamma functions by taking the appropriate limit of $f(z,\gamma)$:
\begin{eqnarray}\label{eq:momrat}
  \psi_{\infty}  & = &
    \lim_{z\rightarrow \infty} \psi(z,\gamma) = \frac{{\rm \dot{M}}_{\rm
      wind}v_{\rm wind}v_{\rm jet}\,h_{1}}{4\pi\,a\,L_{\rm jet,kin}}
      \lim_{z \rightarrow \infty} f(z,\gamma) \nonumber \\
      & = & \frac{{\rm \dot{M}}_{\rm
  wind}v_{\rm wind}v_{\rm jet}\,h_{1}}{4\pi\,a\,L_{\rm jet,kin}}
      \frac{\sqrt{\pi}}{2}\frac{\Gamma(3/2 - 1/\gamma)}{\Gamma(2 - 1/\gamma)}
\end{eqnarray}
where 
\begin{equation}
  f(\gamma) \equiv \lim_{z\rightarrow \infty} f(z,\gamma) =
  \frac{\sqrt{\pi}}{2} \frac{ \Gamma(3/2 - 1/\gamma) }{ \Gamma(2 - 1/\gamma) }
\end{equation}
with $f(4/3) = 1.2$ and  $f(5/3)=1.07$ for a relativistic and non-relativistic monatomic gas,
respectively.  Hence, for otherwise identical parameters, the jet should be bent about 10\%
less in the case of $\gamma=5/3$ compared to the case of $\gamma=4/3$, showing that the effect
of adiabatic index on $\psi_{\infty}$ is moderate.  Figure~\ref{fig:angGam} shows the dependence
of the $f(\gamma)$ on the adiabatic index from $\gamma=4/3$ to $\gamma=5/3$.

While $h_{1}$ cannot be measured observationally, we can express eq.~(\ref{eq:momrat}) in terms
of the observable jet width on VLBA scales, $h_{\rm obs}$ measured at distance $z_{\rm obs}
\gg a$, or alternatively, the observed opening angle $\alpha_{\rm obs}=h_{\rm obs}/(2z_{\rm obs})$:
\begin{eqnarray}
  \psi_{\infty} & = & \frac{h_{\rm obs}}{a}\left(\frac{a^2}{a^2 +
      z_{\rm obs}^2}\right)^{1/\gamma}\frac{{\rm \dot{M}}_{\rm wind}v_{\rm
      wind}v_{\rm jet}}{4\pi L_{\rm jet,kin}}f(\gamma) \nonumber \\
  & \approx & \alpha_{\rm obs} \left(\frac{z_{\rm
        obs}}{a}\right)^{1-2/\gamma}\frac{{\rm \dot{M}}_{\rm wind}v_{\rm
      wind}v_{\rm jet}}{2\pi L_{\rm jet,kin}}f(\gamma)
\end{eqnarray}
where $z_{\rm obs} = z_{\rm VLBA}/\sin(\theta_{\rm LOS})$ and $\alpha_{\rm obs}$ are assumed to be corrected for
foreshortening given the line-of-sight inclination angle $\theta_{\rm LOS}$ of the jet.

\begin{figure}[!htbp]
   \centering
   \includegraphics[width=0.51\textwidth]{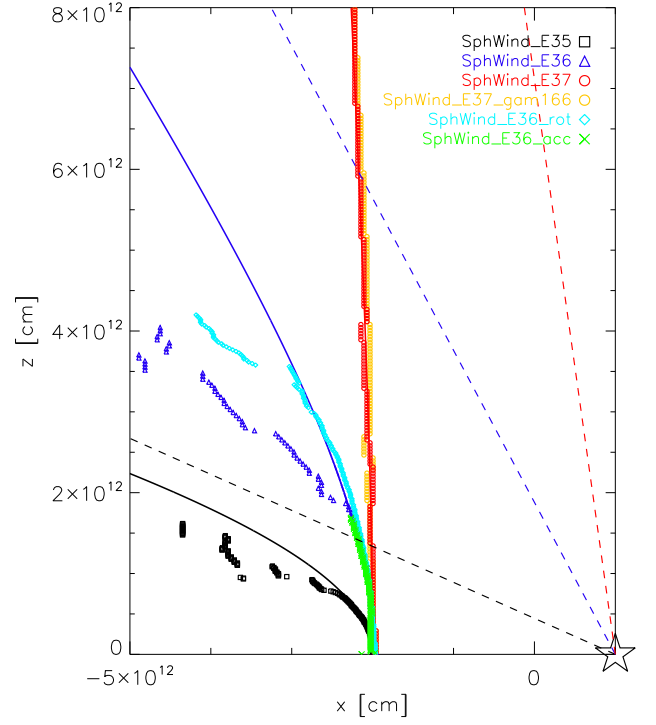}
   \caption{Comparison of numerical results with analytic
     estimates. Each symbol represents the numerical result, and
     solid lines indicate the analytic jet trajectories. Dashed lines
     indicate the asymptotic towards which the jet is expected to
     converge analytically.}
\label{fig:jet_trace}
\end{figure}

We carried out one test simulation, SphWind\_E37\_gam166, with $\gamma=5/3$ for the
jet fluid, and the resulting jet appears more straight, consistent with
analytic expression (see Figure~\ref{fig:jet_trace}).  Note that the ratio is independent of
the orbital separation, $a$, because of the relationship $h_{1} = 2 z_{1} \sin{\alpha_{0}}$, where
$z_{1}$ can be calculated from eq.~(\ref{eq:z1}).

While these expressions are strictly non-relativistic, it is straight forward to show that in
the ultra-relativistic case, the estimate for the bending angle $\psi$ is increased by a factor
of $\sqrt{2}$ over the non-relativistic case (where the velocity is simply set to $v=c$). Thus,
the lower limits we derive below on $L_{\rm jet}$ from observational upper limit on $\psi$
becomes {\em stronger} in the relativistic case.

In the small bending angle regime, our analytic expressions are consistent with the simulation
results, as shown in Figure~\ref{fig:jet_trace}. In the figure, the solid curves were constructed
by integrating the jet trajectory along $z$, given the analytic expression for the transverse
velocity $v_{\rm jet,\perp} \approx \psi v_{\rm jet}$ from eq.~(\ref{eq:psi}) in the small
angle approximation [{\it i.e.,} $d(x-x_{0})/dz=\tan{\psi}$ where $x_{0}$ is the location of
the black hole]. For small bending angles, the figure shows excellent agreement between the
model and the simulations.

As expected, for the case of stronger jet power, $L_{\rm jet}=10^{37}{\rm\, ergs\,s^{-1}}$,
the jet is only moderately bent from the initial direction, while lower power jets are more
strongly affected by the winds, resulting in a higher degree of deflection.  In the figure,
the solid lines represent analytic trajectories calculated by eqs.~(\ref{eq:momJ})-(\ref{eq:momW})
and dashed lines indicate the asymptotic direction estimated in eq.~(\ref{eq:momrat}).

In the case of our fiducial simulation (jet power $L_{\rm jet}=10^{36}{\rm\, ergs\,s^{-1}})$,
the jet becomes dynamically unstable around $z=2.5\times10^{12} {\rm \,cm}$, which leads to
significant broadening of the jet and enhanced bending, and, as a result, the jet begins to
deviate from the analytic estimate.  Further quantitative analysis of the turbulent structures
and their effect to the evolution of the jet is beyond this work. It is, however, clear from
the simulations that jet-instability will only {\em increase} $\Delta \Phi_{m,{\rm wind}} / \Phi_{m,{\rm jet}}$
and $\psi_{\infty}$.

\begin{figure}[!htbp]
   \centering
   \includegraphics[width=0.51\textwidth]{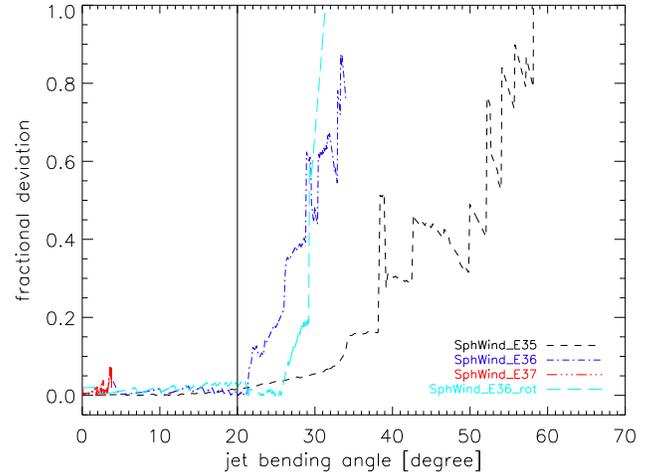}
   \caption{Fractional deviation of the jet trajectory between
     numerical result and analytic formula as a function of jet
     bending angle.}
\label{fig:jetdev}
\end{figure}

Thus, our small jet bending angle approximation can be considered a robust lower limit of
the actual jet bending angle. Figure~\ref{fig:jetdev} shows the fractional deviation of the
numerical result from the analytic estimate as a function of jet bending angle. Our analytic
formula is very accurate for bending angles smaller than 20$^\circ$.  For larger bending
angles, the approximation breaks down.

\subsection{The Effects of Orbital Motion and Wind Acceleration}\label{subsec:Accel}

We carried out two simulations including the effects of orbital motion(SphWind\_E36\_rot)
and radiatively driven winds (SphWind\_E36\_acc) in order to evaluate the importance of both
effects by comparing to our standard model.

Figure~\ref{fig:jet_trace} shows that, for small bending angles, the effects are small and the
results with and without orbital motion and wind acceleration are consistent with each other.
This is not surprising, since the centrifugal force from the orbital motion is about 3 orders
of magnitude less than radiative force, and the Coriolis force acts purely in the transverse
direction.  Therefore, our setup with fixed black hole and star positions is sufficiently
accurate in the context of this analysis.

We reach a similar conclusion about the effect of radiative acceleration. The model with a
radiatively driven wind has a negligibly small difference in the momentum flux of the stellar
wind at the binary separation compared to our standard model.  It implies that our assumption
that the wind reaches terminal velocity before encountering the jet is valid.

For large bending angles, where dynamical instabilities lead to rapid jet disruption and increased
bending, the deviation between individual simulations is noticeable, as expected given the time
variability of the jet trajectory on those scales.

\subsection{Jet Bending as a Diagnostic of Jet Power: The case of Cygnus X-1}\label{subsec:cygx1}

In the limit that $\Delta \Phi_{m,{\rm wind}} > \Phi_{m,{\rm jet}}$, bending in the simulation is so strong
that the jet is dynamically disrupted, rather than simply bent. The interaction disperses the
jet into a broad, no longer collimated flow at much lower velocity than the jet velocity. We
would not expect such a strongly bent jet to survive as an observable radio jet. This suggests
a simple diagnostic: If a stable compact jet is observed intact in an HMXB, one can conclude
that the bending angle should be moderate.

For example, the compact VLBA jet of Cyg X-1 is extended, with a scale length $z_{\rm VLBA}$ of
approximately 15 mas (compared to the angular scale of the orbital separation $a$ of 0.1 mas),
with an upper limit to the half-opening angle of $\alpha_{\rm VLBA} < 2^{\circ}$  where the
viewing angle, $\theta_{\rm LOS}$, is approximated to 40$^\circ$ \citep{Stirling_01}. Combined
with the other fiducial parameters of Cyg X-1, a robust limit can be derived from the observed
stable compact jet by taking the bending angle to be $\psi_{\infty} = \Delta \Phi_{m,{\rm
wind}}/\Phi_{m,{\rm jet}} \ll \pi/2$, which gives
\begin{align}\label{eq:limit}
    & \left( \frac{L_{\rm jet}}{10^{37}\, {\rm ergs\,s^{-1}}} \right)
    \left( \frac{\sin{\theta_{\rm LOS}}}{\sin{40^{\circ}}}\right)^{-2/\gamma} 
    \left( \frac{\alpha_{\rm VLBA}}{2^{\circ}} \right)^{-1} \nonumber \\
    & \times \left[ \frac{f(\gamma=4/3)}{f(\gamma)}\right]  
     \left[\left(\frac{z_{\rm VLBA}}{a}\right) / 150 \right]^{2/\gamma-1}
     \left[ \frac{(150)^{2/\gamma-1}}{(150)^{1/2}}\right] \nonumber \\
    & \times \left( \frac{v_{\rm wind}}{1.6\times10^{8}\,{\rm cm\,s^{-1}}} \right) ^{-1}
      \left( \frac{v_{\rm jet}}{ 0.6\,c} \right)^{-1} \nonumber \\
    & \times \left( \frac{{\rm\dot{M}_{\rm wind}}}{2.6\times 10^{-6}\,{\rm M_{\odot} \,yr^{-1}}} \right)^{-1}   
    \gg 8.5\times10^{-3}
\end{align}
where $\gamma=4/3$ in our standard model. For the case of $\gamma=5/3$, the limit increases
to $3.9\times10^{-2}$ due to relatively shallow increase in jet thickness along the jet, which
requires a larger initial opening angle $\alpha_{0}$ and thus a wider initial jet to give the
same observed $\alpha_{\rm VLBA} \lesssim 2^{\circ}$.

This limit can be made more specific by the fact that the jet appears to be oriented in the
same direction in separate VLBA radio observations, taken during different orbital phases
\citep{Stirling_01}. The data suggest a possible moderate bending on the jet of less then
10$^\circ$ on VLBA scales.  Given the stability of the jet, we consider this an upper limit on
the jet bending angle, {\it i.e.}, $\psi_{\infty} \lesssim 10^{\circ}$, which translates to
\begin{align}\label{eq:limit2}
    L_{\rm jet}  ~ \gtrsim ~ & 7.6\times 10^{35}\,{\rm ergs\,s^{-1}} \nonumber \\ 
       &\times \left( \frac{{\rm\dot{M}_{\rm wind}} } {2.6\times 10^{-6}\,{\rm M_{\odot} \,yr^{-1}}} \right) 
        \left( \frac{v_{\rm wind}}{1.6\times10^{8}\,{\rm cm\,s^{-1}}} \right) \nonumber \\
       &\left( \frac{v_{\rm jet}}{0.6\,c} \right) 
        \left( \frac{\alpha_{\rm VLBA}}{2^{\circ}} \right) 
       \left( \frac{\sin{\theta_{\rm LOS}}}{\sin{40^{\circ}}} \right)^{2/\gamma} 
       \left[\frac{f(\gamma)}{f(\gamma=4/3)}\right] \nonumber \\ 
       &\left[ \left( \frac{z_{\rm VLBA}}{a} \right) / 150 \right]^{1-2/\gamma}
        \left[ \frac{(150)^{1-2/\gamma}}{(150)^{-1/2}}\right],
\end{align} 
where for $\gamma=5/3$, this limit increases to $3.5 \times 10^{36}\,{\rm ergs\,s^{-1}}$.

This lower limit on the jet power is consistent with the range of jet powers quoted in
\citet{Gallo_05}, \citet{Russell_07}, and \citet{Sell_15}, but derived from a completely
independent, likely more robust argument.

\subsection{Off-axis Jets}\label{subsec:offaxis}

When the progenitor of the compact object in an X-ray binary undergoes a supernova explosion
in its last stage of stellar evolution, it likely receives a substantial kick. Such a kick is
capable of leaving the binary in an orbit where the spin and the orbital angular momentum are
misaligned \citep{Brandt_95}.

The spin will slowly align with the orbital axis through a combination of the Lense-Thirring
effect and the internal viscosity of the accretion disk \citep{Bardeen_75}. \citet{Martin_08}
showed that the alignment time scale is a few times smaller than the life time of the mass-transfer
state, typically $t_{align} \sim 10^{6}-10^{8} \rm \, yr$. While this implies that in most XRBs,
spin and orbital angular momentum are likely aligned, it is plausible that they are mis-aligned
in a sub-set of young XRBs, especially for HMXBs, given the short main sequence life time of
the companion.

In this situation, the jet-wind interaction will become phase dependent.  Because the jet
propagation and bending time is short compared to the binary orbital period, the analysis
presented above applies only to the phase of the binary orbit where the jet is perpendicular
to the orbital separation vector $\vec{a}$.  For any given orbit, there are two
such node points.  During other orbital phases, one side of the jet will approach the companion
star ({\it i.e.}, the minimum distance between jet and star is smaller than the orbital separation),
while the other side of the jet is receding; the approaching side will be more strongly bent.

We performed a set of simulations to explore this scenario. We define the inclination angle
$\pi/2 - \theta_{0}$ between the initial direction of the jet $\hat{\vec{z}}$ and the orbital
separation vector $\vec{a}$, such that 
\begin{equation}
  \theta_{0}=\pi/2 - \cos^{-1}{\left(\hat{\vec{z}}\cdot\hat{\vec{a}}\right)}
\end{equation} 
{\it i.e.}, $\theta_{0}=0$ implies the jet is perpendicular to $\vec{a}$, corresponding to the case
discussed so far, while $\theta_{0}=\pi/2$ implies a jet maximally inclined, pointed at the
star/away from it. In other words, in the limit of instantaneous reaction of the jet to orbital
changes, the angle $\phi_{\rm jet,orbit}$ between the jet and the orbital velocity vector does not
affect jet bending (reflected also in the fact that most of our simulations neglect orbital motion
entirely). Clearly, for any non-zero jet inclination relative to the orbital axis, the angle
$\theta_{0}$ will change as a function of orbital phase.

Because jet bending reacts instantaneously to orbital changes, the jet will always propagate 
in a plane spawned by the initial jet direction and the orbital separation vector, to 
order considered here. For circular binary orbits, the asymptotic bending angle therefore
only depends on $\theta_{0}$, and only implicitly depends on binary phase through $\theta_{0}$.

We now extend our analytic formula to the off-axis case. The momentum flux of the jet will
be the same, but the accumulated momentum flux of the stellar wind changes with inclination
angle. The accumulated wind momentum per unit jet length is then (following eq.~(\ref{eq:momW})):
\begin{equation}
  \label{eq:momW2} 
  \Delta \Phi_{m,{\rm wind}}(\theta) = 
  \frac{h_{1}{\rm \dot{M}}_{\rm wind}v_{\rm wind}} {4\pi\,a\,v_{\rm jet}} \, \tilde{f}(z,\gamma,\theta_{0}),
\end{equation}
where
\begin{equation}
    \tilde{f}(z,\gamma,\theta_{0}) \equiv 
    \left( \cos{\theta_{0}} \right)^{-2+1/\gamma} \int^{z/a}_{-\tan{\theta_{0}}} dy \left( \frac{1}{1+y^2} \right)^{2-1/\gamma}.
\end{equation}

We can use the ratio of the accumulated wind momentum to jet momentum per unit length,
eq.~(\ref{eq:momrat}), to derive the asymptotic bending angle $\psi_{\infty}$
\begin{equation}
  \psi_{\infty} =
  \frac{{\rm \dot{M}}_{\rm wind}v_{\rm wind}v_{\rm jet}\,h_{1}}{4\pi\,a\,L_{\rm jet,kin}} \,\tilde{f}(\gamma,\theta_{0}),
\end{equation}
where 
\begin{equation}
   \tilde{f}(\gamma,\theta_{0}) \equiv \lim_{z\rightarrow \infty} \tilde{f}(z,\gamma,\theta_{0}).
\end{equation}
It is obvious that if $\theta_{0}$ is 0, this expression reduces to eq.~(\ref{eq:momrat}).

\begin{figure}[!htbp]
   \centering 
   \includegraphics[width=0.51\textwidth]{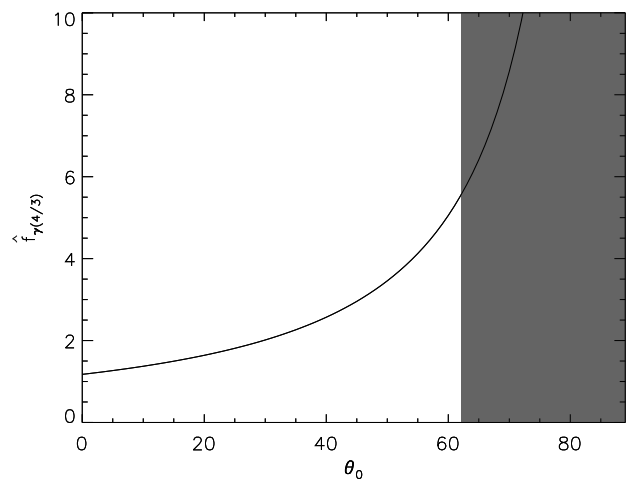} 
   \caption{Variation of $\tilde{f}$
   as a function of $\theta_{0}$. The grey area shows the impact region where the approaching jet runs into the stellar surface.}
\label{fig:c2} \end{figure}

For the approaching side of the jet, which is inclined towards the companion star in our simulation
(Figure~\ref{fig:c2}), $\tilde{f}(\gamma,\theta_{0})$ is monotonically increasing, resulting in stronger bending of the jet.

\begin{figure}[!htbp] 
\begin{center}$ 
  \begin{array}{cc} 
    \includegraphics[width=0.25\textwidth]{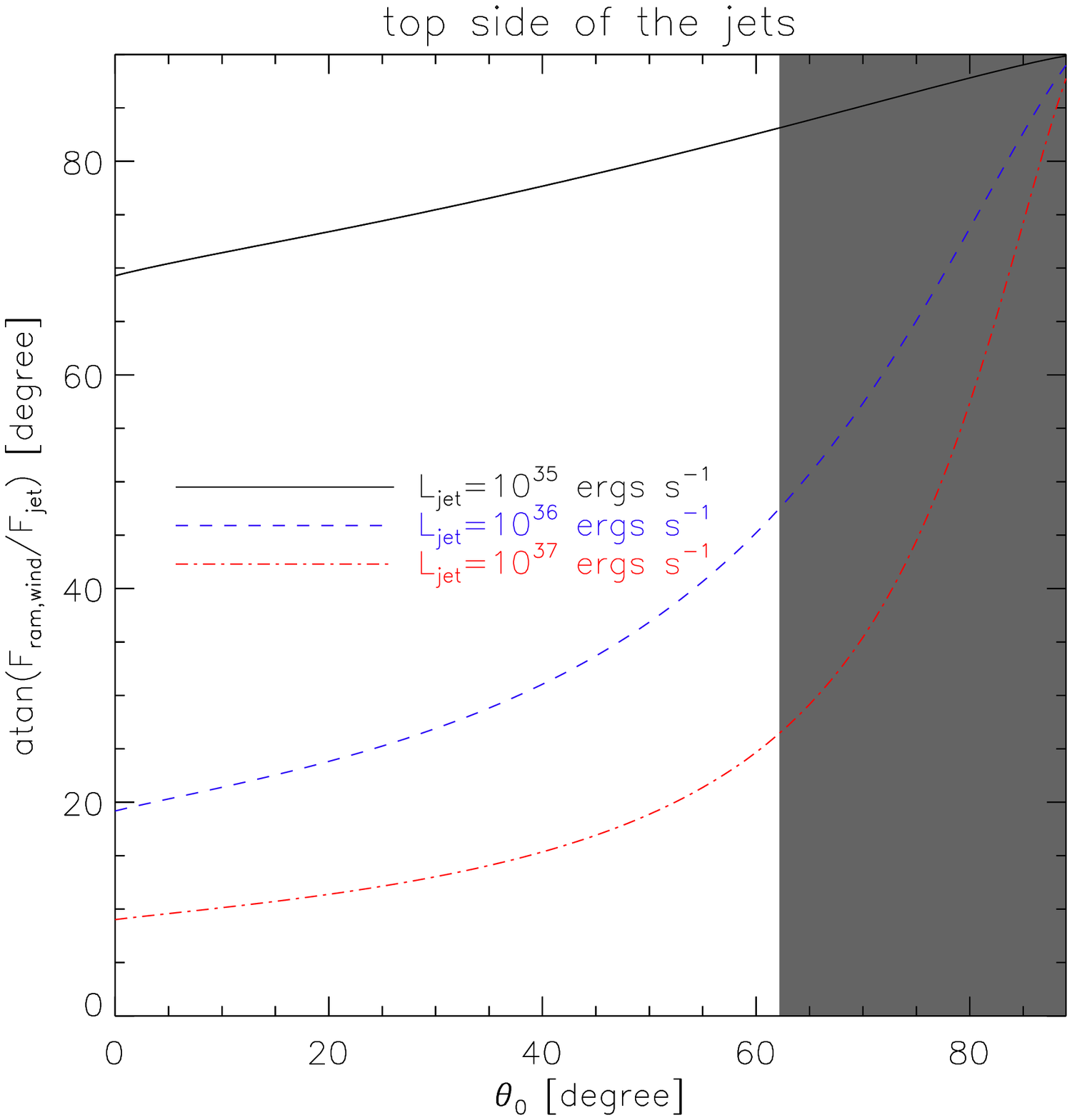} &  
    \includegraphics[width=0.25\textwidth]{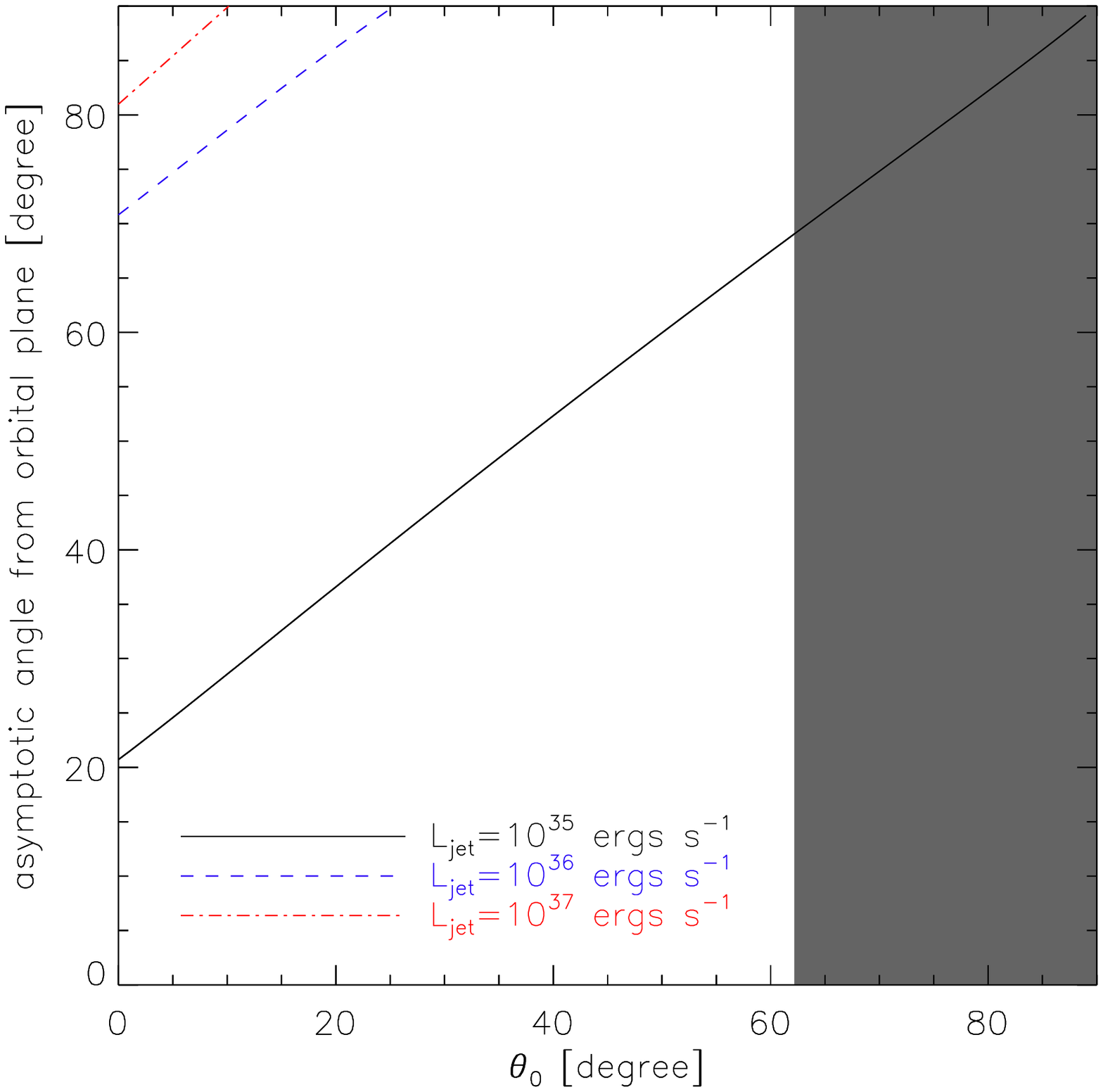}    
  \end{array}$ 
\end{center} 
\caption{Case of approaching jet, initially pointing towards the
  companion star.  Left panel: The ratio of the accumulated wind
  momentum flux to the jet momentum flux as a function of
  $\theta_{0}$.  Right panel: Asymptotic jet bending angle with
  respect to the orbital plane. The grey area shows the impact region
  where the approaching jet runs into the stellar surface.}
\label{fig:angrat} 
\end{figure}

If $\theta_{0}$ is larger than some value, the jet collides with the star.  Given the
binary parameters for Cygnus X-1, this limit angle is 62.18$^\circ$. The impact
region where the jet directly encounters the star is marked by the grey area in the
Figure~\ref{fig:c2}\&\ref{fig:angrat}.

Figure~\ref{fig:angrat} shows $\psi_{\infty}$ as a function of $\theta_{0}$, assuming $h_{1}$ is
kept constant in the approaching side of the jet where it heads toward the companion star. From
this analysis we can infer the jet bending angle for an initially off-axis jet. As one might
expect, a more inclined jet will be more strongly bent.  The right panel of Figure~\ref{fig:angrat}
shows $\psi_{\infty}$ with respect to the orbital plane. The figure shows a monotonic increase
of jet bending angle with initial jet inclination.

\begin{figure}[!htbp] 
\begin{center}$ 
  \begin{array}{cc} 
    \includegraphics[width=0.25\textwidth]{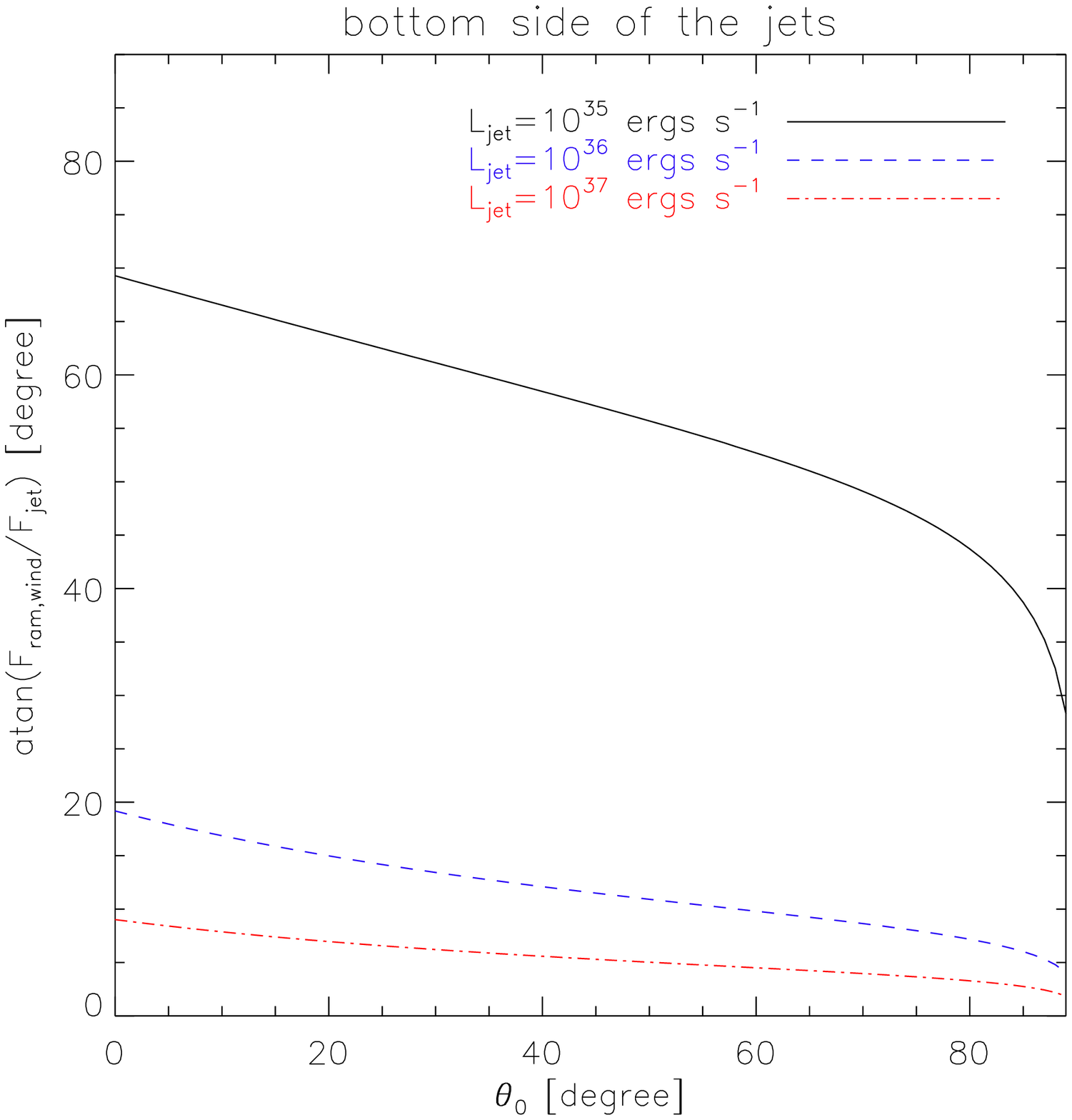} &  
    \includegraphics[width=0.25\textwidth]{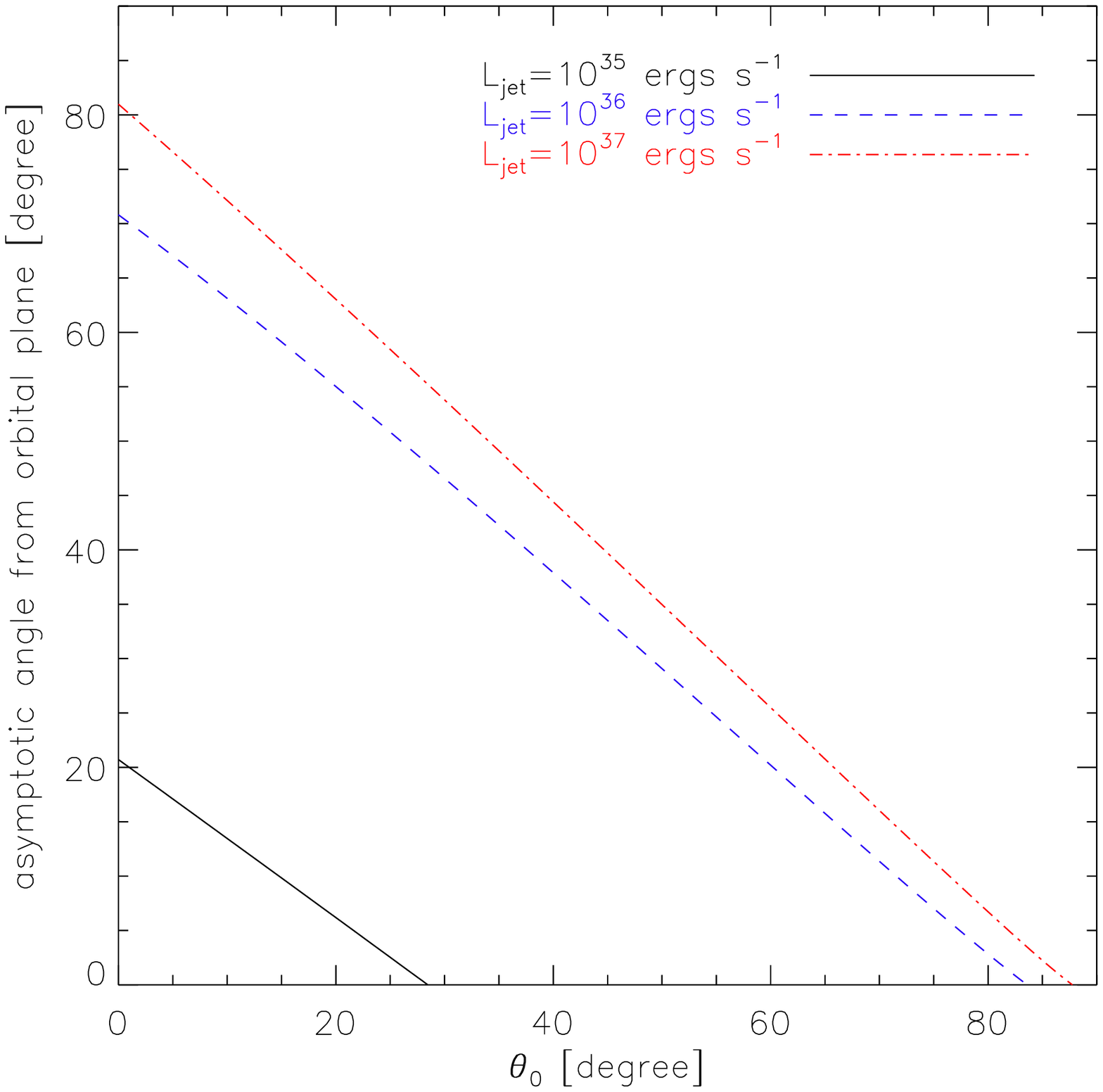}    
  \end{array}$ 
\end{center} 
\caption{Case of receding jet, initially pointing away from the
  companion star. Same panels as in Figure~\ref{fig:angrat}}
\label{fig:angrat_bt} 
\end{figure}

Figure~\ref{fig:angrat_bt} shows the momentum ratio and asymptotic angle for the receding jet.  For
jets with higher inclination angles, the bending angle is reduced.

\begin{figure*} 
   \centering
   \includegraphics[width=0.95\textwidth]{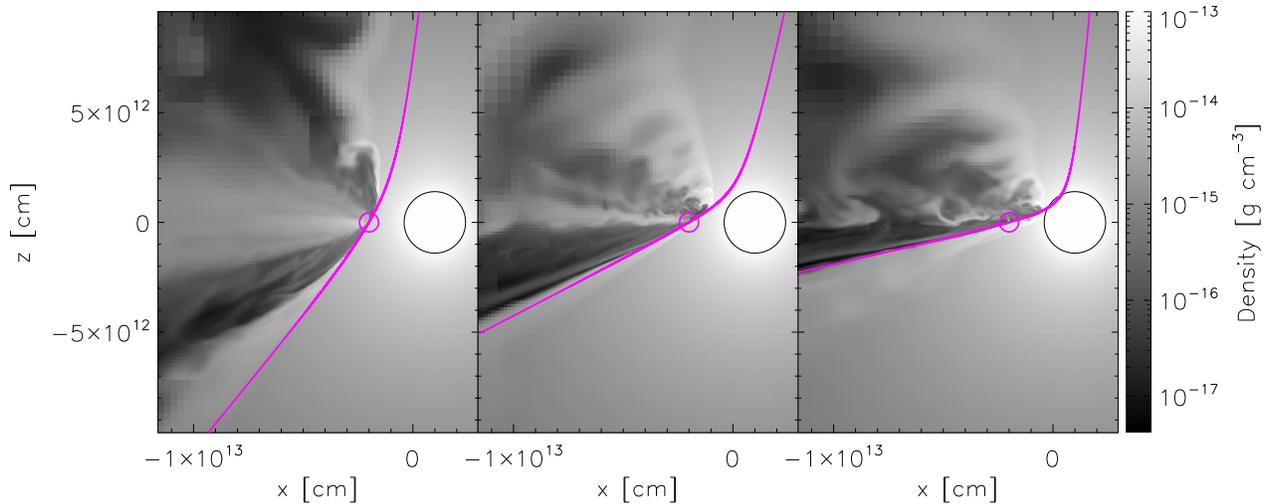}
   \caption{Density maps for off-axis jets, deviating from
  perpendicular direction to orbital plane:
  $30^{\circ},~60^{\circ},~75^{\circ}$ from left to right panel. The
  solid magenta line indicates the analytical trajectory of the
  jet. The magenta circle indicate the location of the black hole.}
\label{fig:1e36_deg}
\end{figure*}

Figure~\ref{fig:1e36_deg} shows density maps for several cases of the misaligned simulations.
All other parameters are the same as in our standard model, SphWind\_E36. The inclination angles
are $30^{\circ},\,60^{\circ},\,75\,^{\circ}$, respectively.  The magenta solid line represents the analytic
jet trajectory which is derived using the assumption of pressure balance between the jet and
the ambient medium.  For all cases, the approaching jet shows some degree of disruption and
dynamical instability and the analytic approximation breaks down for bending angles larger than
about 20$^\circ$. On the other hand, the figures show that our analytic approach is acceptable
for the receding side of the jets.

As one would expect from simple geometric considerations, while a jet launched perpendicular
to the orbital separation $\vec{a}$ is bent symmetrically on either side, an inclined jet shows
asymmetric behavior between approaching jet and receding jet. Such a configuration
would lead to increased jet bending in the approaching jet.  Our limit on the jet power in Cyg
X-1 was derived under the most conservative assumption that the jet is not inclined relative to
the orbital axis. Because jet bending and disruption increase for inclined jets, the possible
inclination of the jet will strengthen our limit on the jet power.

\citet{Szostek_07} argued that in Cygnus X-1, the orbital modulation of radio emission occurs
due to free-free absorption in the asymmetric wind as a function of orbital phase, and the
observed phase lag of the modulation with respect to the orbital phase could be attributed to
the time delay for the emission from the dynamically curved jet.  However, the jet bending we
studied in this work is likely not the cause of the phase lag, because the bending direction
is parallel to the orbital separation vector, leading to only very small phase lags due to
light-travel-time delays, while the observed phase lags would require bending in the direction
against the orbital velocity, as argued by \citet{Szostek_07}.

For higher jet inclination angles, ({\it e.g.}  SphWind\_E36\_60deg or SphWind\_E36\_75deg),
both jet and counter-jet will impact the star once per orbit, leading to jet disruption and
reformation; one might thus expect to observe episodic jet eruption from such a system.

\subsection{The case of Cygnus X-3}

We can apply a similar argument to the case of Cygnus X-3, an HMXB consisting of a compact object
and a Wolf-Rayet companion.  In this case, it is still not clear whether the compact object is a
black hole or a neutron star, but the observational evidence of collimated jet-like structures is
strong in multi-epoch radio maps \citep{Marti_01}.  The radio jets with $\gamma$-ray flares occur
during bright soft X-ray states or during state transitions \citep{Szostek_08, Tavani_09}.

\citet{Dubus_10} suggested that the jet of Cygnus X-3 is inclined relative to the orbital plane
($20^{\circ} < \psi_{\rm jet} < 80^{\circ}$) in order to obtain good fits to the gamma-ray modulation,
with a significant offset between the jet footpoint and the site of the gamma-ray emission.

This estimate does not take potential jet bending into account.  In fact, as shown above, even
if the orbital angular momentum and the jet axis are aligned, we should expect some bending to
occur, and the asymptotic jet direction will change over the course of the binary orbit.  As a
result, the precession of the jet occurs on a time scale, $\tau_{\rm precess}\approx\tau_{\rm
orbit}$ in our model. We should point out that there is currently no evidence
for (or against) jet precession in Cygnus X-3. The observed gamma-ray modulation can be
explained by an inclined jet. An inclined jet is likely bent more than a jet perpendicular to
the orbital plane, so jet bending by the jet-wind interaction may have to be taken into account
in detailed models of Cygnus X-3.  

Assuming the initial jet direction is perpendicular to the orbital plane, we can estimate
the bending angle from eq.~(\ref{eq:momrat}) from the binary and jet parameters $L_{\rm
jet}=10^{38}\,{\rm ergs\,s^{-1}}$, $\dot{\rm M}_{\rm wind}=10^{-5}\, {\rm M_{\odot}\,yr^{-1}}$,
$a=3\times10^{11} {\rm cm}$, $v_{\rm wind}=10^{8}\,{\rm cm\,s^{-1}}$, and $v_{\rm jet}=0.5\,c$ as used by
\citet{Dubus_10}.

Note that Cygnus X-3 is a considerably tighter system with higher inferred jet energy than
Cygnus X-1, while the momentum flux from its neighboring Wolf-Rayet star is higher than that
from OB companion in Cygnus X-1.  In terms of the fiducial parameters given above, the jet
bending angle can be calculated from eqs.~(\ref{eq:z1}) and (\ref{eq:momrat}) as
\begin{align}
  \psi_{\infty} \sim & \,2.35^{\circ}  \nonumber \\
  & \times \left( \frac{{\rm\dot{M}_{\rm wind}} } {10^{-5}\,{\rm M_{\odot} \,yr^{-1}}} \right)^{1/2}
  \left( \frac{v_{\rm wind} }{ 10^{8}\,{\rm cm\,s^{-1}}} \right)^{1/2} 
  \left( \frac{v_{\rm jet} }{ 0.5\,c} \right)^{1/2} \nonumber \\ 
  &\left( \frac{L_{\rm jet}} {10^{38}\,{\rm ergs\,s^{-1}}} \right)^{-1/2} 
   \left[\frac{f(\gamma)}{f(\gamma=4/3)}\right]
  \left[ \frac{\sin{\alpha_{0}}}{\sin\left( 2^{\circ} \right)} \right] ,
\end{align}
indicating that for jet bending to be important, the initial opening angle of the jet would have
to be significantly larger than a few degrees, or the jet would have to be strongly misaligned with
the orbital axis, with the jets very closely approaching the stellar surface for part of the orbit.

\subsection{Caveats}
\label{sec:caveats}
Our simulations were carried out using an isotropic and homogeneous stellar wind model. This
assumption allows us to study the dynamics of jet-wind interaction in a simple analytic
framework. Several complications will affect this process in ways not included in this paper.

\subsubsection{X-ray Ionization}
Firstly, the X-ray flux from the accretion disk may be intense enough to ionize the circum-stellar
gas, so the wind can be reduced or quenched, decreasing the jet bending angle. Line driving
is likely inefficient in the illuminated portion of the wind if the ionization parameter is
above a critical threshold. Conversely, the X-ray illumination can itself produce thermal wind
driving by heat input. The exact configuration of the wind is therefore not entirely clear.

However, from the fact that the black hole in Cygnus X-1 is accreting from the companion wind,
the illuminated portion of the wind must be of similar density to that assumed in the shadowed
region, which suggests that the estimates presented above will not be changed drastically by
the effects of ionization of the wind.

\subsubsection{Clumping}
More importantly, the stellar wind from OB-type stars is likely clumpy
\citep{Oskinova_12}. \citet{Poutanen_08} suggested that the wind in Cygnus X-1 is clumped,
based on dips in the X-ray lightcurve.

This situation was simulated by P12.  Because global simulations that incorporate clumped winds
with realistic clump sizes and densities will be computationally impossible in the foreseeable
future, given the large dynamic range and resolution required to capture the mass in small
clumps, our paper should be considered complementary to the work presented in P12.

While a detailed treatment of clumping in our simulations is beyond
the scope of this paper, we will briefly discuss how significant
clumping would affect the results of our simulations.  For a fixed
mass loss rate and wind velocity, the net effect of wind clumping will
depend on the filling factor, the average size of the clumps, and the
density contrast.  Since secular bending is facilitated by the
pressure gradient across the jet, we should expect bending in the
presence of clumps to be reduced by the amount the momentum flux in
the hot low-density background component of the wind is reduced
relative to an un-clumped wind.  The reduction is set by the density
ratio of the hot/low density background component $\rho_{\rm hot}$ to
the density $\rho_{\rm wind}$ of an un-clumped wind of the same
emission measure.

Since $\rho_{\rm hot}$ is poorly constrained, we cannot easily
quantify the reduction expected in the bending angle by the presence
of clumps.  Clearly, if $\rho_{\rm hot}$ is of the same order as the
estimates of $\rho_{\rm wind}$ used above, our results will be
unaffected. However, if $\rho_{\rm hot} \ll \rho_{\rm wind}$, the
amount of secular bending of the jet will be reduced by the density
ratio $\rho_{\rm hot}/\rho_{\rm wind}$.

In the case of O-star winds in HMXBs, the presence of clumps in the
illuminated side of the wind is even less well constrained, given the
effects X-ray ionization may have on clump formation.

For all these reasons, a general statement about how our results
change in the presence of clumping is difficult to make. However,
considering the possible limits of clump sizes and density, we can
estimate under which circumstances the jet would be able to propagate
through the wind without disruption and derive similar constraints on
the jet power for a given set of clump parameters.

If the density contrast between the clumps and the background wind is
small, corrections should naturally be minor and the limit on the jet
power from eq.~(\ref{eq:limit2}) will hold.  If the density contrast
is large and most of the mass is carried in clumps, the effect will
depend on the average size of the clumps, relative to the width $h$ of
the jet:

{{\bf Large clumps:} If the clumps are large compared to the cross
  section of the jet, jet-clump interaction will be disruptive, and the conclusions of P12
  hold, {\it i.e.}, the jet will be disrupted unless it is sufficiently powerful to escape,
  in the case investigated by P12, $L_{\rm jet} \gtrsim 10^{37} \,{\rm
    ergs\,s^{-1}}$, similar to the conclusion we reach in this paper,
  with the main difference being that interaction with a clumpy wind introduces significant
  stochasticity. It has been suggested by \citet{Zdziarski_11} that fluctuations in the radio
  light curve may be an expression of such interaction.}

{{\bf Small clumps:} If the clumps are much smaller than the cross
  section of the jet, such that the typical clump radius satisfies
  $R_{\rm clump} \ll h_{\rm jet}$, they will act like bullets passing
  through the jet, generating small bow shocks that slow down a small
  fraction of the jet. Small clumps will only globally disrupt the jet
  if {\bf (a)} the covering fraction $f_{\rm clumps}$ of
  clumps is larger than unity {\em and} {\bf (b)} the total mass
  within clumps is sufficiently high to stop the jet fluid. {\em Both}
  conditions must be met for the jet to be disrupted by a wind whose
  mass flux is dominated by small clumps.
  
  Note that in this case, jet disruption will not be stochastic, since
  it requires the presence of many small clouds passing through the
  jet continuously, rather than a few large clouds. Since the jet in
  Cygnus X-1 is {\em not} disrupted continuously, at least one of the
  two conditions for jet disruption must be violated. We can now
  estimate the constraints the observed stability of the jet against
  wind disruption place on the jet and cloud parameters.}
  
  \begin{itemize}
  \item[{\bf (a)}]{{\bf Low covering fraction:} The covering fraction
      ${f}_{\rm cover}$ of clumps is given by
      \begin{equation}
        f_{\rm cover} = \int_{0}^{\infty} dz \frac{f}{R_{\rm clump}} \sim
        \frac{a f_{\rm vol}}{R_{\rm clump}}
      \end{equation}
      where $z$ is the path length along the jet, $a$ is the orbital
      separation, and $f_{\rm vol}$ is the clump volume filling
      factor.  If ${f}_{\rm cover} \gg 1$, the jet {\em
        may} be disrupted if the total mass within clumps is
      sufficiently high to stop the jet fluid (${{f}}_{\rm
        clumps} \gg 1$ is a necessary condition for jet disruption).
      Conversely, a {\em sufficient but not necessary} condition for
      the jet {\em not} to be disrupted is that the covering fraction
      of clumps is small, ${f}_{\rm cover} \ll 1$, or
      \begin{equation}
        \label{eq:coveringfraction}
        f_{\rm vol} \ll \frac{R_{\rm clump}}{a} \ll 1
      \end{equation}

      The condition that the emission measure of the clumped wind is
      the same as that of the un-clumped wind (in the optically thin
      limit) gives a clump volume filling fraction of $f_{\rm vol} =
      \rho_{\rm wind}^2/\rho_{\rm clump}^2$ or $\sqrt{f_{\rm
          vol}}\,\rho_{\rm clump}/\rho_{\rm wind} = 1$.  With
      eq.~(\ref{eq:coveringfraction}), the condition on 
      ${{f}}_{\rm cover}$ implies that the wind mass loss
      rate must satisfy
      \begin{eqnarray}
          {\rm \dot{M}_{\rm wind,clumped}} &=& \frac{\rho_{\rm clump}
          f_{\rm vol}}{\rho_{\rm wind}}\dot{\rm M}_{\rm wind} 
          = \sqrt{f_{\rm vol}}\,\dot{\rm M}_{\rm wind} \nonumber \\
          &\ll& \dot{\rm M}_{\rm
          wind}\sqrt{\frac{R_{\rm clump}}{a}} \ll \dot{\rm M}_{\rm wind}
        \label{eq:condition_1}
      \end{eqnarray}}
  \item[{\bf (b)}]{{\bf Low mass density:} A second necessary
      condition for clumps in the wind to significantly disrupt the
      jet is that the clump mass $\rm \Delta M_{\rm clump}$ intercepted by
      the jet exceed the inertial mass of the jet plasma, $\rm M_{\rm
        jet}$. Conversely, a second {\em sufficient but not necessary}
      condition for the jet {\em not} to be disrupted is
    \begin{eqnarray}
        {\rm M_{\rm jet}} &=& \int_{0}^{\infty}dz A_{\rm jet} \rho_{\rm jet,inertial} \nonumber \\
                          &\gg& \int_{0}^{\infty} dz A_{\rm jet} f \rho_{\rm clump} = \Delta {\rm M_{\rm clump}}
    \end{eqnarray}
    which we can simplify as the condition
    \begin{equation}
      \rho_{\rm jet} \gg f_{\rm vol}\rho_{\rm clump}
    \end{equation}
    
    Assuming that the location of the re-collimation shock of the jet
    is set by the ram pressure of the hot, low density background gas
    of the clumped wind, we can use eqs.~(\ref{eq:P_jetRam}) and
    (\ref{eq:P_windRam}) to write this condition as
    \begin{equation}
      f_{\rm vol}\frac{\rho_{\rm clump}}{\rho_{\rm hot}} \ll  \frac{v_{\rm wind}^2
        {\mathcal M}_{\rm jet,0}^2}{v_{\rm jet}^2} = \frac{v_{\rm wind}^2}{c_{\rm s,jet,0}^2}
    \end{equation}
    where $\rho_{\rm hot} \ll \rho_{\rm wind} \ll \rho_{\rm clump}$ is
    the density of the hot background wind in the clumped wind
    scenario. This condition can be written as a limit on the wind
    mass flux:
    \begin{equation}
      {\rm \dot{M}_{\rm wind,clumped}} \ll \frac{\rho_{\rm hot}}{\rho_{\rm wind}}
      \frac{v_{\rm wind}^2}{c_{\rm s,jet,0}^2} {\rm \dot{M}_{\rm wind}} \ll
      {\rm \dot{M}_{\rm wind}}
      \label{eq:condition_2}
    \end{equation}
    where the last inequality reasonably assumes that $c_{\rm s,jet,0}
    > v_{\rm wind}$.}
\end{itemize}

In summary, we can distinguish the following three cases, one of which
must apply to Cygnus X-1:
\begin{enumerate}
\item{The wind is not strongly clumped (that is, the mass flux in the
    diffuse/background wind is comparable to the mass flux used in this paper,
    even if clumps are present). In this case, our limit in
    eq.~(\ref{eq:limit2}) applies and the jet must be powerful to
    propagate through the wind.}
\item{The wind is strongly clumped, with large clumps of size $R
    \gtrsim h$; in this case, the analysis of P12 applies and the jet
    must be powerful to propagate through the wind.}
\item{The wind is strongly clumped, with small clumps of size $R \ll
    h$; in this case, the mass loss rate must be orders of magnitude
    below the mass loss rate of an un-clumped wind, given the observed
    emission measure and lack of jet disruption. In this case, we
    cannot use the observed lack of jet disruption/bending to derive a
    limit on the jet power.}
\end{enumerate}
Thus, (a) either the wind mass loss rate must be orders of magnitude
smaller than the nominal wind parameters for a uniform wind for the
companion star, with important implications for wind formation in
HMXBs, and/or (b) the jet power must be large [eq.~(\ref{eq:limit2})],
similar to the conclusion reached from energy estimates of the large
scale nebula.

\section{Conclusion}

We performed global hydrodynamic simulations to study jet-wind interaction in the wind of the
companion star in HMXBs. The interaction results in the jet being bent with a characteristic
bending angle that depends on the properties of the jet and the wind.  Using the small bending
angle approximation, we derived a simple analytic formula for the asymptotic bending angle. The
formula is consistent with the numerical results, and it can be used to analyze observations
of jets in HMXBs. We showed that the analysis is valid for bending angles smaller than about
20$^\circ$.

We applied the formula to two observed HMXBs, Cygnus X-1 and Cygnus X-3.  We constrained
the jet power in Cygnus X-1 to be $L_{\rm jet} \gtrsim 10^{36} \,{\rm
ergs\,s^{-1}}$ from the lack of observed precession of
the VLBA jet. This limit is consistent with previous estimates, but derived from completely
independent arguments and likely more robust.

Given Cygnus X-3 parameters, we argued that the jet bending is likely not significant, unless
the jet in Cygnus X-3 has a large opening angle $\alpha$ or is significantly less powerful than
estimated in previous studies.

The main caveat in the application of this model to observed HMXBs is the lack of knowledge
about the properties of wind clumping in massive stars. We discussed conditions under which
clumping will affect the analysis of jet bending presented in this paper, and under which
clumps themselves will lead to observable dynamic disruption of the jet. We showed that, if
wind clumping is dynamically important for the jet-wind interaction in Cygnus X-1, our limit on
the jet power holds unless the wind mass loss rate is orders magnitude below the nominal wind
parameters derived for a uniform wind (in which case we would not be able to constrain the jet
power from the lack of disruption or bending by the wind).

\acknowledgments{We acknowledge grant support through NSF award AST-0908690. We would like to thank
Rob Fender, Rashid Sunyaev, and Enrico Ramirez-Ruiz for helpful discussions.  We thank
Andrzej Zdziarski for insightful and very helpful comments and suggestions that helped improve
the paper significantly.}

\bibliographystyle{apj}
\bibliography{xrbs_sphWind}

\end{document}